\documentclass[sigconf,screen]{acmart}

\AtBeginDocument{%
  \providecommand\BibTeX{{%
    \normalfont B\kern-0.5em{\scshape i\kern-0.25em b}\kern-0.8em\TeX}}}

\setcopyright{rightsretained}
\acmPrice{}
\acmDOI{10.1145/3597926.3598088}
\acmYear{2023}
\copyrightyear{2023}
\acmSubmissionID{issta23main-p168-p}
\acmISBN{979-8-4007-0221-1/23/07}
\acmConference[ISSTA '23]{Proceedings of the 32nd ACM SIGSOFT International Symposium on Software Testing and Analysis}{July 17--21, 2023}{Seattle, WA, United States}
\acmBooktitle{Proceedings of the 32nd ACM SIGSOFT International Symposium on Software Testing and Analysis (ISSTA '23), July 17--21, 2023, Seattle, WA, United States}
\received{2023-02-16}
\received[accepted]{2023-05-03}

\usepackage{graphicx}
\usepackage{subcaption}
\usepackage{xcolor}
\usepackage{xspace}
\usepackage{multirow}
\usepackage{adjustbox}
\usepackage{array}
\usepackage{pifont}
\usepackage{enumitem}
\usepackage[bottom]{footmisc}
\usepackage{xstring}
\usepackage{bm}
\usepackage{algorithm}
\usepackage[noend]{algorithmic}
\usepackage[normalem]{ulem}
\usepackage{microtype}
\usepackage{listings}
\definecolor{mygreen}{rgb}{0,0.6,0}
\definecolor{mygray}{rgb}{0.5,0.5,0.5}
\definecolor{mymauve}{rgb}{0.58,0,0.82}

\newcommand{\code}[1]{\lstinline[keywords={}]{#1}}

\newcommand*{\figu}{{Figure}\xspace}

\newcommand*{\tabl}{{Table}\xspace}

\newcommand*{\algo}{{Algorithm}\xspace}

\newcommand{\listitem}[1]{\noindent\textbf{#1}\xspace}

\newcommand{\tool}{\textsc{ACETest}\xspace}
\newcommand{\toolx}{\textsc{ACETest*}\xspace}

\usepackage[most]{tcolorbox}

\def\BibTeX{{\rm B\kern-.05em{\sc i\kern-.025em b}\kern-.08em
    T\kern-.1667em\lower.7ex\hbox{E}\kern-.125emX}}

\begin{document}

\title{\tool: Automated Constraint Extraction for Testing Deep Learning Operators}




\author{Jingyi Shi}
\affiliation{
    \institution{Institute of Information Engineering, CAS; School of Cyber Security, UCAS}
    \city{Beijing}
    \country{China}
}
\additionalaffiliation{
  \institution{Key Laboratory of Network Assessment Technology, CAS}
}
\additionalaffiliation{
  \institution{Beijing Key Laboratory of Network Security and Protection Technology}
}
\email{shijingyi@iie.ac.cn}

\author{Yang Xiao}
\affiliation{
    \institution{Institute of Information Engineering, CAS; School of Cyber Security, UCAS}
    \city{Beijing}
    \country{China}
}
\authornotemark[1]
\authornotemark[2]
\authornote{Corresponding author.}
\email{xiaoyang@iie.ac.cn}

\author{Yuekang Li}
\affiliation{
    \institution{University of New South Wales}
    \city{Sydney}
    \country{Australia}
}
\email{yli044@e.ntu.edu.sg}

\author{Yeting Li}
\affiliation{
    \institution{Institute of Information Engineering, CAS; School of Cyber Security, UCAS}
    \city{Beijing}
    \country{China}
}
\authornotemark[1]
\authornotemark[2]
\email{liyeting@iie.ac.cn}

\author{Dongsong Yu}
\affiliation{
    \institution{Zhongguancun Laboratory}
    \city{Beijing}
    \country{China}
}
\email{yudongsong@zgclab.edu.cn}

\author{Chendong Yu}
\affiliation{
    \institution{Institute of Information Engineering, CAS; School of Cyber Security, UCAS}
    \city{Beijing}
    \country{China}
}
\authornotemark[1]
\authornotemark[2]
\email{yuchendong@iie.ac.cn}

\author{Hui Su}
\affiliation{
    \institution{Institute of Information Engineering, CAS; School of Cyber Security, UCAS}
    \city{Beijing}
    \country{China}
}
\authornotemark[1]
\authornotemark[2]
\email{rhondasu928@gmail.com}

\author{Yufeng Chen}
\affiliation{
    \institution{Institute of Information Engineering, CAS; School of Cyber Security, UCAS}
    \city{Beijing}
    \country{China}
}
\authornotemark[1]
\authornotemark[2]
\email{chenyufeng@iie.ac.cn}

\author{Wei Huo}
\affiliation{
    \institution{Institute of Information Engineering, CAS; School of Cyber Security, UCAS}
    \city{Beijing}
    \country{China}
}
\authornotemark[1]
\authornotemark[2]
\email{huowei@iie.ac.cn}



\begin{abstract}

Deep learning (DL) applications are prevalent nowadays as they can help with multiple tasks.
DL libraries are essential for building DL applications.
Furthermore, DL operators are the important building blocks of the DL libraries, that compute the multi-dimensional data (tensors).
Therefore, bugs in DL operators can have great impacts.
Testing is a practical approach for detecting bugs in DL operators.
In order to test DL operators effectively, it is essential that the test cases pass the input validity check and are able to reach the core function logic of the operators.
Hence, extracting the input validation constraints is required for generating high-quality test cases.
Existing techniques rely on either human effort or documentation of DL library APIs to extract the constraints.
They cannot extract complex constraints and the extracted constraints may differ from the actual code implementation.

To address the challenge, we propose \tool{}, a technique to automatically extract input validation constraints from the code to build valid yet diverse test cases which can effectively unveil bugs in the core function logic of DL operators.
For this purpose, \tool{} can automatically identify the input validation code in DL operators, extract the related constraints and generate test cases according to the constraints.
The experimental results on popular DL libraries, TensorFlow and PyTorch, demonstrate that \tool{} can extract constraints with higher quality than state-of-the-art (SOTA) techniques.
Moreover, \tool{} is capable of extracting 96.4\% more constraints and detecting 1.95 to 55 times more bugs than SOTA techniques.
In total, we have used \tool{} to detect 108 previously unknown bugs on TensorFlow and PyTorch, with 87 of them confirmed by the developers.
Lastly, five of the bugs were assigned with CVE IDs due to their security impacts.

\end{abstract}




\maketitle

\section{Introduction}


Deep learning (DL) has been applied in almost every inch of modern society, including computer vision~\cite{CV1, CV2}, natural language processing~\cite{BERT, NLP2}, and program comprehension~\cite{Ahmad2020ATA,PatrickEvans2021XFLNF,Liu2019ATOMCM}, etc. 
As the backbone of DL systems, DL models are used to process the multi-dimensional inputs, which are called tensors.
To develop, deploy and execute DL models with reusable code, data scientists and engineers have developed various DL libraries such as TensorFlow~\cite{TensorFlow} and PyTorch~\cite{Paszke2019PyTorchAI}.
Inside the DL libraries, DL operators are the key components to perform computational tasks on tensors and one library API may use multiple DL operators.
Since the logic of the DL operators is computational intensive, the DL operators are implemented as low-level components of the libraries with C/C++ programming language in order to execute faster.
As a result, bugs in these DL operators are often memory errors with security impacts and can affect the functionality of multiple APIs in the DL library.
Unfortunately, DL operators are vulnerable to bugs.
For example, TensorFlow, one of the most popular DL libraries, has disclosed 378 vulnerabilities in the CVE database since 2019~\cite{cve-mitre}.
Furthermore, 249 (65.87\%) of them are related to DL operators.
Therefore, early detection of bugs in DL operators is important for assuring the quality of DL libraries and applications.

Testing~\cite{Sen2005CUTEAC,Wang2020DeepLL,Cadar2011SymbolicEF,She2018NEUZZEF,Wang2017SkyfireDS} is a popular technique for detecting bugs in programs.
Compared with static analysis techniques~\cite{Liu2021DetectingMS,Xiao2020MVPDV,Cai2022PeahenFA,Fan2019SMOKESP,Wang2017AutomaticDA}, it produces much fewer false positives and provides proof-of-concept test cases for human experts to reproduce the bugs.
However, generating high-quality test cases for DL operators is challenging.
The main reason is that most of the bugs in DL operators reside within the core functional logic responsible for processing tensors, while reaching this logic requires the test cases to pass a set of validity checks.
Randomly generated test cases can hardly pass these validity checks, and passing the validity checks requires the test cases to satisfy the constraints of the input validation logic.
Therefore, acquiring the constraints for valid inputs is important for generating high-quality test cases.

Several techniques have been proposed to acquire the input constraints for DL operators or DL library APIs in recent years.
Both Duo~\cite{Duo} and Predoo~\cite{Predoo} require to supply the input constraints manually when detecting bugs in DL operators.
The drawback of manual input constraints extraction is two-fold.
On the one hand, it requires a lot of manual effort.
On the other hand, some input constraints are too complex to be extracted manually.
To save manual effort, FreeFuzz~\cite{Wei2022FreeLF} proposes to mine test cases from open source and DeepREL~\cite{DeepREL} leverages existing test cases to test more APIs by inferring relational APIs. IvySyn ~\cite{ivysyn} directly tests the C/C++ native code of DL operators by instrumenting to obtain type constraints of input parameters.
However, due to the lack of clarity regarding constraints when generating test cases, they can not generate test cases effectively.
To automatically extract complex constraints, DocTer~\cite{DocTer} uses rule-based approaches to collect constraints from the API function descriptions of the documentation.
Nevertheless, DocTer has difficulties in extracting input constraints effectively for DL operators.
The quality of the extracted constraints heavily relies on the API function descriptions.
If the API function description is absent or inaccurate, the extracted constraints will be affected.
Moreover, the API function description may not fully reflect the logic of the actual implementation.
Therefore, an approach to automatically extract the input constraints accurately for DL operators is needed.

Symbolic Execution (SE)~\cite{king1976symbolic,Li2013SteeringSE,dudina2017using} is a widely used approach to extract constraints by traversing all paths in the source code.
However, it suffers from scalability problems because of the infamous path explosion problem.
Additionally, the constraint solving can be time-consuming and challenging for large programs~\cite{Zhang2020MultiplexSE}.
Directed Symbolic Execution (DSE)~\cite{DART} can partially mitigate the scalability problems by extracting constraints related to specific paths instead of all the paths in the program.
However, the problem of whichever specific paths to traverse is left untouched.
If the paths are not selected properly, the collected constraints can still cause scalability issues.
Hence, directly applying existing SE or DSE techniques cannot fulfill the need.

To fill the research gap, we propose \tool, a technique which can automatically extract input validation constraints from the code of DL operators and use the constraints to generate high-quality test cases for detecting bugs in DL operators.
The overall idea of \tool is to extract constraints related to the user-controllable inputs for input validation and leverage them to generate the test cases randomly within the valid input space.
Thus, the generated test case is valid enough to reach the core functional logic, and is diverse enough to find bugs.
\tool works in three steps: input validation path extraction, constraint extraction and testing.
Initially, \tool gathers meta information of the DL operators, including their names and parameter types.
Next, by traversing backwards on the control-flow graph (CFG) from the error-handling basic-blocks (BBs), \tool identifies the input validation code.
With the input validation code, \tool extracts all execution paths that do not result in errors during input validation.
For each path, \tool extracts constraints related to the user-controllable inputs. To accomplish this, a general constraint model is proposed to guide the constraint extraction process.
The tool employs controllability propagation rules to identify user-controllable values in a restricted manner during the constraint extraction process. Additionally, a set of constraint construction rules is utilized to construct constraints while traversing the path.
Once the constraints are extracted, the tool employs Z3~\cite{Z3} to generate solutions and subsequently generates test cases based on these solutions. The generated test cases are then utilized to test the DL operators, and any resulting crashes are reported accordingly.

We implemented \tool{} as an automatic DL operator testing framework and evaluated its performance on 610 and 485 operator-level APIs for TensorFlow and PyTorch, respectively.
The experiment results demonstrate that \tool can extract 96.4\% more constraints than DocTer.
With the extracted constraints, the generated test cases of \tool can increase the pass rate on TensorFlow and PyTorch by 162.60\% and 138.58\% than DocTer, respectively. 
Consequently, \tool detects 1.95 to 55 times more bugs than SOTA techniques.
Notably, \tool is able to detect 108 previously unknown bugs, with 87 already confirmed by the developers, and five of them received CVE numbers.

In summary, our paper makes the following contributions:
\vspace{-3pt}
\begin{itemize}[leftmargin=*]
    \item 
    \listitem{Dimension.}
    This paper opens a new dimension for generating high-quality test cases for DL operators by automatically extracting input validation constraints from the code.
    \item
    \listitem{Technique.}
    We build \tool{}, a DL operator testing framework that can extract input validation constraints automatically and generate high-quality test cases.
    \tool{} can automatically identify the input validation paths and extract the constraints related to user inputs from the code of the DL operators.
    \item
    \listitem{Evaluation and Impact.}
    With experiments conducted on operator-level APIs of TensorFlow and PyTorch, we found that the constraints extracted by \tool{} exhibit a higher quality than those extracted by existing SOTA techniques.
    Furthermore, \tool{} detected 108 previously unknown bugs with 87 of them confirmed by the developers.
    Additionally, five of the bugs were assigned CVE IDs.
\end{itemize}

The source code of \tool{} as well as the list of identified bugs will be publicly accessible at \url{https://github.com/shijy16/ACETest}.





\section{Background}

\subsection{DL Library Testing}\label{sec:bg_DLLibTest}

Depending on the test targets, existing DL library testing techniques fall into three categories: model-related testing, high-level API testing, and operator testing.
In \textit{model-related testing}, DL models are generated as test cases and executed using the target DL library. This form of testing focuses on evaluating various aspects of the library's functionality, such as model parsing logic, computational graph optimization logic, execution scheduling logic, and other aspects of model handling. While multiple components within the DL library are tested, it is often challenging to reach the deep logic of each component.
This difficulty arises from the lack of clear linkage between the features of the models and the features of the underlying code, making it hard to generate models that can specifically exercise certain code components.
\textit{High-level API testing} focuses on the high-level APIs in DL libraries.
By generating test cases for each high-level API directly, these techniques enjoy more flexibility than model-related testing and they can test deeper into each API function. 
\textit{Operator testing} further narrows the scope to the operators in DL libraries.
Operator testing can be conducted by invoking the API for each operator (namely, operator-level API), which is typically a lower-level approach compared to testing high-level APIs.
Operators are the key components for carrying out computational tasks.
They are the essential building blocks for the DL library APIs as well as model handling logic.
Moreover, operators are often implemented with C/C++, meaning that they are vulnerable to memory errors which are security critical.

\vspace{-5pt}
\subsection{Running Example}

\begin{figure}[t]
	\centering
	\includegraphics[width=1.0\linewidth]{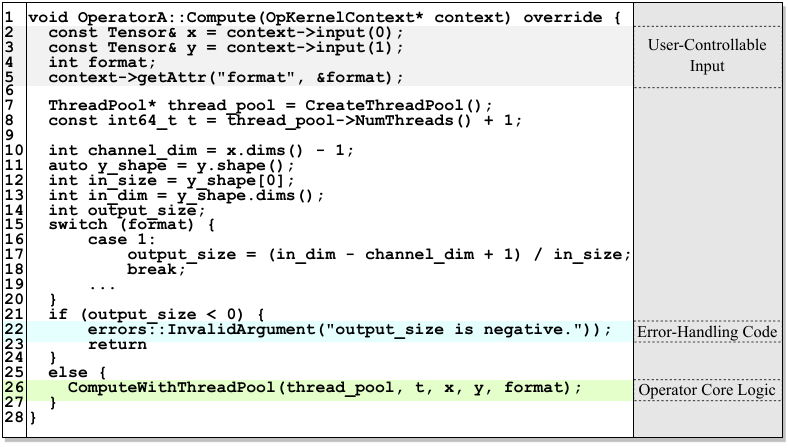}
         \vspace{-18pt}
	\caption{A running example}
	\label{fig:running}
        \vspace{-5pt}
\end{figure}

\figu{}~\ref{fig:running} shows the source code of \code{OperatorA} which is a running example crafted by modifying the operator \code{Conv2D} from TensorFlow.
A bug exists in the \code{ComputeWithThreadPool} function (line~26 in \figu{}~\ref{fig:running}).
From \figu{}~\ref{fig:running}, we can obtain the following observations step-by-step.
\ding{182} Although symbolic execution can be used for solving path constraints and exploring various program states, the buggy function can be too complex to handle for symbolic execution.
In this example, \code{ComputeWithThreadPool} is using a thread pool  which introduces a tremendous amount of program states.
So, symbolic execution is very likely to encounter the path explosion problem here.
Therefore, \textit{testing is a more practical approach} to detect bugs in DL operators.
\ding{183} In order to reach the buggy function, the test inputs must fulfill several input validation constraints so that the value of \code{output_size} is larger than or equal to zero.
Therefore, \textit{fulfilling input validation constraints is important} for generating meaningful test inputs to test the core logic of the DL operators.
In this example, the constraints are relatively complex and they are not described in the documentation.
Hence, an approach to \textit{extract constraints from the source code} is needed.
\ding{184} To extract the input validation constraints, we need to identify the input validation code.
From the example, we can observe that the input validation code contains error handling logic to deal with invalid inputs (line~22 in \figu{}~\ref{fig:running}).
Thus, \textit{error handling logic can help to locate the input validation code}.
\ding{185} Not every line of code before the core logic is related to input validation.
In this example, the call to the \code{CreateThreadPool} function is not relevant with input validation, despite its placement within the input validation code.
Therefore, we need to \textit{exclude irrelevant code} when extracting input validation constraints.

\vspace{-5pt}
\section{Approach}

\begin{figure*}[t]
	\centering
	\includegraphics[width=1.0\textwidth]{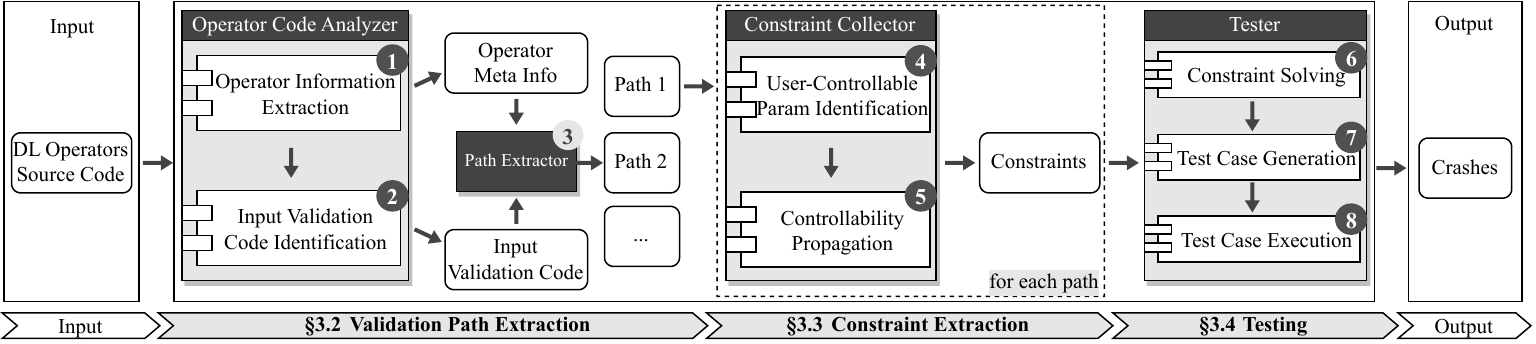}
	\caption{Overview of \tool}
	\label{fig:overview}
\end{figure*}

\subsection{Overview}

\figu{}~\ref{fig:overview} shows the workflow of \tool{}.
The overall input is the source code of the target DL operators and the overall output is a set of test inputs which can crash the DL operators (crashes).
The workflow contains three main steps:
validation path extraction, constraint extraction and testing.

\listitem{Input Validation Path Extraction.}
\ding{182} \tool{} extracts the basic information of the DL operators from the source code of the DL library, including the information shown in \tabl{} ~\ref{table:meta-info}.
\ding{183}
\tool{} identifies the input validation logic of the DL operators according to the extracted information in the previous step.
\ding{184}
\tool{} extracts the execution paths of the input validation code.

\listitem{Constraint Extraction.}
\ding{185} For every extracted input validation path, \tool{} first identifies the user controllable inputs which can serve as the entrances for testing.
\ding{186} Then, \tool{} propagates the user controllability along the path and extract the input constraints at the same time.
After this step, \tool{} can get the constraints for each extracted input validation path.

\listitem{Testing.}
\ding{187} Given a set of constraints for an input validation path, \tool{} uses Z3 to get the solutions.
\ding{188} \tool{} generates test cases by filling up the constraint solutions and randomly generating the rest.
This ensures that the test cases can both pass the input validation logic and have enough diversity to explore the core functional logic.
\ding{189} \tool{} executes the target DL operator with the test cases and observes for crashes.

\begin{table}[t]
  \caption{Operator Meta Information}
    \label{table:meta-info}
    \centering
    \vspace{-8pt}
    \begin{adjustbox}{max width=\linewidth}
    \begin{tabular}{l|l|l}
            \hline
            \textbf{Item Name} & \textbf{Example (Running Example)} & \textbf{Used By} \\ \hline
            Operator Name & \code{OperatorA} & \ding{182}
            \\ \hline
            Operator-level API Name & \code{tf.raw\_ops.OperatorA} & \ding{189}
            \\ \hline
            Parameter List & \code{x}, \code{y}, \code{format} & \ding{185}, \ding{188}
            \\ \hline
            Parameter Type & \code{tensor}, \code{tensor}, \code{int} & \ding{186}, \ding{188}
            \\ \hline
            Source Code Location & \textit{tensorflow/core/kernels/OperatorA.cc} & \ding{183}
            \\ \hline
    \end{tabular}
    \end{adjustbox}
\end{table}







\subsection{Validation Path Extraction}

\subsubsection{Operator Information Extraction}
\tabl{}~\ref{table:meta-info} shows the meta information of operators that \tool{} extracts from the code.
\textit{Operator name} is needed in this step for extracting all other information.
\textit{Operator-level API name} is needed for invoking the DL operators for testing.
\textit{Parameter list} is useful for locating the user controllable inputs in the source code of operators and generating test cases.
\textit{Parameter types} are needed for collecting the constraints and solving them.
\textit{Source code locations} of the DL operators are used for extracting the input validation logic.





\begin{figure}[t]
	\centering
	\includegraphics[width=1.0\linewidth]{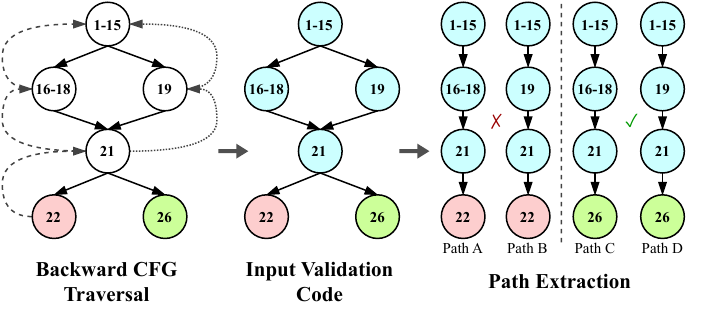}
	\caption{Input validation path extraction in the running example. Line 22 is the error-handling function call, while line 26 represents the core functional logic of the operator. \tool identifies the input validation code by performing a backward traversal from line 22, marking all blue nodes as input validation code. \tool then extracts the execution paths within the input validation code and its direct successors. Paths that reach the core functional logic (line 26) are saved, and in this example, Path C and D are saved.}
    \label{fig:path-extract}
    \vspace{-2pt}
\end{figure}

\subsubsection{Input Validation Code Identification}
Identifying the input validation code is an important step for extracting validation paths.
In conventional software, distinguishing between input validation logic and core functional logic automatically can be challenging, as they are often intertwined.
However, DL operators possess two distinct characteristics which enable the automatic extraction of input validation logic.
\ding{182} Input validation is always conducted before the core functional logic in DL operators. 
DL operators in popular DL libraries (such as TensorFlow and PyTorch) usually support multiple devices such as Intel CPU, Nvidia GPU, AMD GPU, and TPU. 
To optimize hardware resource utilization, these DL operators implement device-specific core functional code. 
Consequently, input validation is uniformly performed prior to entering separate core functional code tailored for different devices.
\ding{183} In case of input validation failure, specific error handling functions will be invoked.
Since DL operators are usually executed in parallel as part of DL models, a global execution context is employed to manage multiple processes.
Error handling functions are used across all operators to notify the context about validation failures.

Based on the two characteristics, \tool{} effectively identifies the input validation code by performing a backward trace from the error handling function calls on the control flow graph (CFG).
Specifically, \tool{} employs a Depth-First Search approach to traverse the CFG and pinpoint the input validation code.
\figu{}~\ref{fig:path-extract} shows how \tool{} utilizes the backward traversal method to identify the input validation code in the running example.
To address the potential issue of path explosion associated with backward traversal, \tool{} sets a threshold to limit the max number of iterations for the search process. The default value for this threshold is set to 1 million.
To mitigate the risk of missing basic blocks when the fixed threshold is reached, we propose an online algorithm that dynamically determines whether a basic-block should be considered as input validation code or not.
This algorithm is invoked only when the threshold is reached, as the DFS process can already find all the basic-blocks associated with input validation.
The rationale of this algorithm is to perform a quick and approximate check of reachability from a given basic-block to the error-handling function calls by examining whether any of the error-handling function calls are in the successors of the given basic-block.
If at least one error-handling function call can be reached from the basic-block, it is considered part of the input validation code.
The detailed algorithm is shown in \algo{}~\ref{alg:test_reach}.
In \algo{}~\ref{alg:test_reach}, $current\_bb$ is the basic-block under examination, and is served as a parameter.
Additionally, $reachable\_cache$, $unreachable\_cache$ and $eh\_bbs$ are global variables.
 $reachable_cache$ and $unreachable_cache$ are lists that respectively store the basic blocks marked as reachable and unreachable, providing a performance boost to the algorithm.
$eh\_bbs$ represents the list of error-handling basic-blocks.



\begin{algorithm}[t] 
	\caption{Basic-block reachability check.} 
	\label{alg:test_reach} 
	\renewcommand{\algorithmicrequire}{\textbf{Input:}}
	\renewcommand{\algorithmicensure}{\textbf{Output:}}
	\begin{algorithmic}[1]
		\REQUIRE $current\_bb$, $reachable\_cache$, $unreachable\_cache$, $eh\_bbs$
		\IF{$reachable\_cache$.\textsc{find}($current\_bb$)}
		    \RETURN \TRUE
		\ENDIF
		\IF{$unreachable\_cache$.\textsc{find}($current\_bb$)}
		    \RETURN \FALSE
		\ENDIF
		\STATE $worklist \leftarrow [{current\_bb}]$
		\STATE $visited \leftarrow [{current\_bb}]$
		\WHILE{$!worklist.$\textsc{empty()}}
		    \STATE $bb \leftarrow worklist.pop()$
		    \IF{$eh\_bbs.$\textsc{find}($bb$) \OR $reachable\_cache$.\textsc{find}($bb$)}
		        \STATE $reachable\_cache$.\textsc{insert}($current\_bb$)
		        \RETURN \TRUE
		    \ENDIF
		    \FOR{$succ : bb.$\textsc{successors()}}
		        \IF{$!visited.$\textsc{find}($succ$)}
		            \STATE $visited.$\textsc{insert}($succ$)
		            \STATE $worklist.$\textsc{insert}($succ$)
		        \ENDIF
		    \ENDFOR
		\ENDWHILE
		\FOR{$bb : visited$}
		    \STATE $unreachable\_cache$.\textsc{insert}($bb$)
		\ENDFOR
		\RETURN \FALSE
	\end{algorithmic} 
\end{algorithm}


\subsubsection{Path Extraction}
After identifying the basic-blocks of the input validation code, \tool{} extracts the execution paths that will be used for constraint extraction in the next stage.
Technically, \tool{} can extract two types of paths from the input validation code and its successors.
The first type of paths end up calling the error-handling functions while the second type of paths lead to the core functional logic.
In practice, the first type of paths never reach the core functional logic, which is generally much more buggy than the input validation logic and the error handling logic. Furthermore, missing/wrong validation checks can also lead to crashes in the functional logic.
Therefore, \tool{} ejects the first type of paths and retains only the second type.
For the running example, as shown in \figu{}~\ref{fig:path-extract}, \textit{Path C} and \textit{Path D} are kept while \textit{Path A} and \textit{Path B} are ejected.


\vspace{-10pt}
\subsection{Constraint Extraction}
\begin{figure*}[t]
	\centering
	\includegraphics[width=1.0\textwidth]{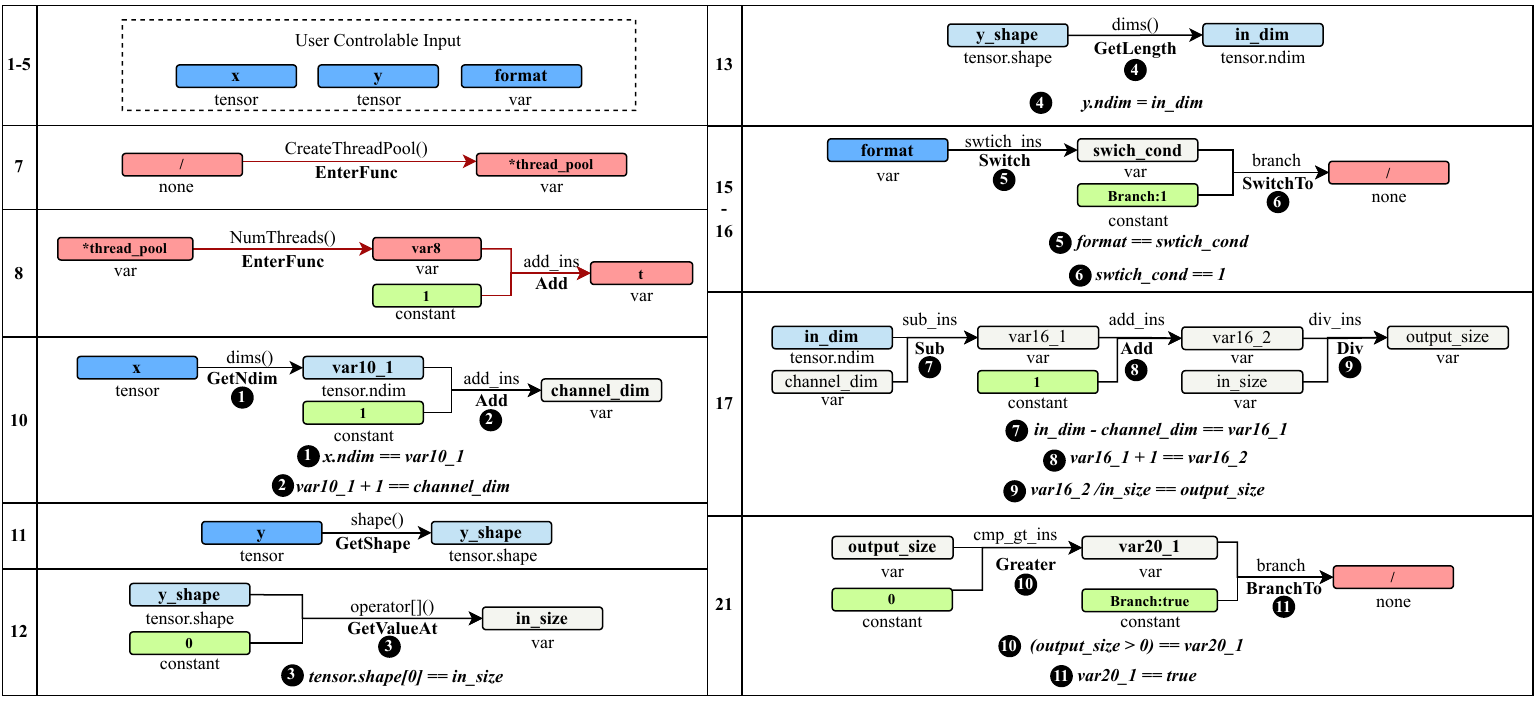}
	\caption{Extracting constraints from a path in the running example. The column 1 and 3 show the line number for the propagation on its right.
	Each node represents a value in the path, the color represents its value type: dark blue for parameters, light blue for properties, gray for other controllable values, green for constants, red for non-controllable values. Black arrow means a propagation goes from left to right, red arrows  means non-propagation. Above each arrow is a current function/instruction, and bellow the arrow is the matched propagation rule. The constraints constructed when propagating are labeled with the same number as the propagation rule.}
	\label{fig:cons_extract}
\end{figure*}

This stage aims at extracting constraints from the paths extracted in the previous stage. Symbolic execution is a common solution for extracting constraints from programs. However, complex structures such as tensor are common in DL operators. The overhead of blindly symbolizing all parameters in DL operators would be unacceptable. Thus, to extract constraints from the given paths in DL operators in a light-weight manner, we propose a novel constraint extraction technique for DL operators. The technique identifies user-controllable parameters(\ref{param_ident}) at first and collects constraints for them by traversing the path from the first instruction to the last instruction. During the traversal process, other user-controllable values are identified through controllability propagation(\ref{propagation}). A constraint model and a set of propagation rules are introduced to restrict the propagation, minimizing the overhead and extraneous constraints. Furthermore, constraint construction rules are applied in the propagation process to construct constraints for user-controllable values based on each instruction.

\subsubsection{User-Controllable Parameter Identification}
\label{param_ident}
User-controllable parameters play a crucial role in constraint extraction. Before extracting constraints in paths, we have to identify all of the user-controllable parameters in DL operators.
User-controllable parameters are passed to operators in different ways across different DL libraries, including direct transmission via operator function parameters or through dedicated function calls that retrieve parameters from the execution context.
For the first kind, identifying user-controllable parameters is straightforward. And the second kind requires manual identification and annotation of relevant functions within the target DL library. The user-controllable parameters are then recognized by identifying calls to these functions.

For Example, PyTorch passes user-controllable parameters directly through the parameters of the operator functions, while TensorFlow through specific function calls. 


\subsubsection{Controllability Propagation}
\label{propagation}
 We traverse the path and propagate the controllability from the values of user-controllable parameters. Constraints are extracted in the propagation process. But propagating blindly to every value polluted by controllable values can result in increasing the overhead of constraint extraction, collecting redundant constraints, and thus burdening constraint solving. Hence, a controllability propagation technique is proposed to address the problem. The technique consists of constraint model, propagation rules, and constraint construction rules. The constraint model helps to reduce the propagation scope. It provides a foundation for the application of propagation rules, which dictate the propagation behavior of each instruction within the given path. Simultaneously, the constraint construction rules are employed to generate corresponding constraints.

\noindent \textbf{Constraint model.}
The basic idea of controllability propagation is to focus only on values and instructions that might be related with constraints. According to our knowledge, most constraints for tensors only constrain their properties such as shape and number of dimensions. Values of elements in tensors are not constrained in most cases. Thus, the propagation can be restricted by focusing on specific properties of each parameter. 
Based on the above observation, we propose a general constraint model for DL operator parameters to guide the propagation process. The model specifies the properties that are commonly constrained for each type of parameter, as shown in \tabl ~\ref{table:properties}.

\begin{table}[t]
\caption{Constraint model}
    \label{table:properties}
    \vspace{-5pt}
     \renewcommand\arraystretch{1.35}
    \centering
    \begin{adjustbox}{max width=\linewidth}
    \small
    \begin{tabular}{c|c|c|p{0.3\textwidth}}
            \hline
            \textbf{Argument} & \multirow{2}{*}{\textbf{Property}} & \textbf{Property} & \multirow{2}{*}{\textbf{Description}} \\ 
            \textbf{Type}&&\textbf{Type}&\\
            \hline
            \multirow{4}{*}{tensor} & $dtype$ & dtype & \multicolumn{1}{m{5cm}}{The data type of the tensor.} \\
            \cline{2-4}
            ~ & $ndim$ & int & \multicolumn{1}{m{5cm}}{The number of dims of the tensor, also the length of the tensor's shape.} \\ 
            \cline{2-4}
            ~ & $num\_element$ & int & \multicolumn{1}{m{5cm}}{The total number of elements in the tensor.} \\ 
            \cline{2-4}
            ~ & $shape[]$ & list of int & \multicolumn{1}{m{5cm}}{The shape of the tensor.} \\
            \hline
            \multirow{4}{*}{list} & $element\_type$ & type & \multicolumn{1}{m{5cm}}{The type of the list's elements.} \\
            \cline{2-4}
            ~ & $length$ & int & \multicolumn{1}{m{5cm}}{The length of the list.} \\
            \cline{2-4}
            ~ & $iter$ & int & \multicolumn{1}{m{5cm}}{The iterator value of the list, a general representation of all the values in the list.} \\
            \cline{2-4}
            ~ & $value[]$ & list of $element\_type$ & \multicolumn{1}{m{5cm}}{The concrete value of the list.} \\
            \hline
            \multirow{2}{*}{string} & $value$ & string & \multicolumn{1}{m{5cm}}{The value of itself.} \\
            \cline{2-4}
            ~ & $format\_id$ & int & \multicolumn{1}{m{5cm}}{The format id of this string if it is a format string that specifies the dimension order of input tensors.} \\
            \hline
            int & $value$ & int & \multicolumn{1}{m{5cm}}{The value of itself.}\\
            \hline
            float & $value$ & float & \multicolumn{1}{m{5cm}}{The value of itself.}\\
            \hline
            bool & $value$ & bool & \multicolumn{1}{m{5cm}}{The value of itself.}\\
            \hline
    \end{tabular}
    \end{adjustbox}
    \vspace{-3pt}
\end{table}

\noindent \textbf{Propagation rules.}
With the help of constraint model, we construct a set of propagation rules to identify properties included in the model and other intermediate values related to constraints. As each instruction is traversed, we match that instruction to one of the propagation rules and perform the corresponding propagation behavior.
There are three kinds of propagation rules: (1) parameter to property; (2) property to property; (3) general propagation. 

\begin{figure}[t]
        \setlength{\abovecaptionskip}{0pt}
	\centering
	\includegraphics[width=0.45\textwidth]{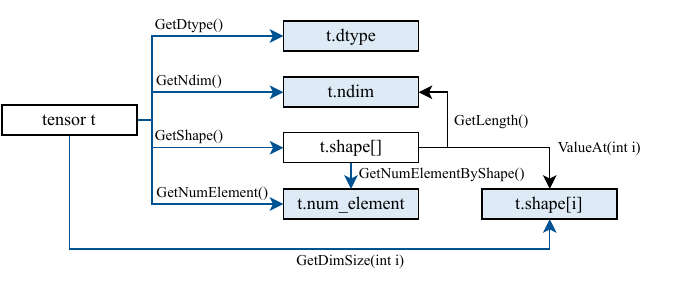}
	\caption{Propagation rules for tensors}
	\label{fig:propagation}
    \vspace{-10pt}
\end{figure}

Parameter to property rules define the propagation behaviour from a parameter to one of its properties. The blue arrows in \figu ~\ref{fig:propagation} show parameter to property rules for tensors. For example, $ s = t.GetShape() $ is a behaviour propagating the controllability from parameter $tensor\ t$ to $s$, and $s$ is then bound to $t$'s property $shape$. 

Properties of a parameter can propagate from one to another in specific conditions. The black arrows in \figu ~\ref{fig:propagation} show the property to property rules for tensor. For example, the behavior expressed by $ n = t.shape.GetLength() $ propagates the controllability from property $shape$ of $tensor\ t$ to another property $ndim$, and $n$ is then bound to $t$'s property $ndim$. 

The first two kinds of rules are involved with transformation between parameters and properties, and transformation between properties themselves. Function calls to specific functions that get properties from parameters or properties are mapped into behaviours defined in these rules. For instance, calls to $Tensor::shape()$ in TensorFlow are mapped to $GetShape$. 
On the other hand, general propagation rules are employed to handle propagation in instructions taht do not involve such transformations, such as numerical calculation instructions and addressing instructions. The basic idea of general propagation rules is to propagate to the output values of an instruction only when all input values are controllable or constant. The general propagation rules are constructed for all types of instructions in LLVM bit-code.

Utilizing the three kinds of rules, we conduct a traversal from the initial instruction to the final instruction along each path. During the traversal, we propagate the controllability from the user-controllable parameters to other values in a restricted manner based on the rules. The first two kinds of rules help us identify the values representing properties defined in the constraint model. And the third kind of rules propagate facilitates the propagation of controllability to other values related to these properties. 

\noindent \textbf{Constraint construction rules.}
The rules of constraint construction composites of two parts: natural constraint construction and propagation constraint construction. 

Natural constraint construction rules are used to restrict the properties of different parameters based on the nature of them. These constraints must be fulfilled by all parameters, otherwise it would contradict the construction of a valid parameter. For example, a tensor's shape length must equal its number of dims.
Additionally, to make sure the size of generated parameters are acceptable, we add rules to limit the size of tensor and list in this step. Natural constraints construction rules are applied to each parameter before the path traversal.

Propagation constraint construction rules are applied during the propagation process. As controllability is propagated from the input values to the output values of an instruction, a constraint is constructed to capture the semantics of the instruction. This process involves mapping each type of instruction to its corresponding rule. 
Furthermore, to improve the accuracy and efficiency of constraint construction, we manually implement construction process for common functions frequently used in input validation. This allows for the skipping of the propagation and constraint construction process within these functions.

\subsubsection{Constraint Extraction Example}
\figu ~\ref{fig:cons_extract} shows how \tool extracts constraints from path C (\figu ~\ref{fig:path-extract}) in the running example.  The detail process is described as follows:
\begin{itemize}[leftmargin=*]
\setlength{\itemsep}{0pt}
\setlength{\parsep}{0pt}
\setlength{\parskip}{0pt}
    \item \textbf{Line 1-5:} The controllable values $tensor\ x$, $tensor\ y$ and $int\ format$, are identified by matching the called functions $input$ and $GetAttr$. In the meantime, \tool dose not enter $input$ and $GetAttr$ as they are already handled.
    \item \textbf{Line 7:} Function $CreateThreadPool$ is called without any controllable inputs, but \tool still enters its body to check for constant return values or potential calls to specific functions that could provide input. But nor of the mentioned situations happens, thus $*thread\_pool$ is not controllable.
    \item \textbf{Line 8:} Similarly, function $NumThreads$ is traversed but its return value $var8$ is not controllable. Thus, the instruction $add$ does not match the propagation rule for it. As the rule requires all its parameters to be controllable or constant, but one of its parameter is not controllable or constant. Thus, its result $t$ is not controllable, and no constraints are constructed.
    \item \textbf{Line 10:} The function $Tensor::dims()$ called here is matched with propagation rule $GetNdim$ for tensors, and its implicit argument is tensor $x$. Thus the return value of this function call, $var10\_1$, is propagated as controllable, and the constraint construction rule bound to $GetNdim$ is also triggered, resulting in the construction of a corresponding constraint (\ding{182}). And then, the propagation rule for the $Add$ instruction is fulfilled because its parameters are either constant or controllable. Thus $channel\_dim$ is propagated to be controllable and a corresponding constraint construction rule for $Add$ is triggered (\ding{183}).
    \item \textbf{Rest of the path:} In the rest of the path, propagation rules and constraint construction rules are applied in the same manner for each instruction. Finally, a total of 11 constraints are constructed as a result.
\end{itemize}

\subsection{Testing}\label{subsec:testing}


\subsubsection{Constraints Solving} 
The constraints are stored as SMT formulas after constraint extraction.
Thus, \tool{} can leverage Z3 for sampling random solutions from the complex SMT formulas with bit-vectors, arrays, and uninterpreted functions. 
As Z3 does not have a built-in batch sampling function, in order to obtain diverse solutions, we continuously insert new constraints to block the newly obtained solution.
After this step, \tool{} can get a batch of unique solutions for the constraints.

\subsubsection{Test Case Generation} 
\tool{} uses the constraint solutions for the constrained properties and randomly generates the remaining unconstrained properties. To generate additional edge cases, \tool{} mutates the dimension sizes of tensors, int and float values by assigning them extremely large values with a probability $p$.
Moreover, \tool{} wraps the test inputs with a Python script which invokes the operator-level APIs , serving as the final test code for execution.




\subsubsection{Test Case Execution} 
We observe that DL libraries suffer a lot from the cold boot overhead (about 1-5 seconds) for each run. 
Thus, \tool maintains a test process which keeps acquiring and executing the final test code from a queue until it crashes.
After a crash, \tool{} will restart the test process and continue testing.
Since there are much more normal test cases than crashing ones, the test process can greatly increase the testing speed of \tool{} by allowing it to execute up to 300 test cases per second.

\section{Implementation}


We implemented \tool in about 5K lines of C++ code and 3K lines of Python code. Input validation path extraction and constraint extraction are implemented as a compiler pass based on LLVM~\cite{LLVM}. We employed  SVF~\cite{SVF} to generate Value Flow Graphs (VFGs) and CFGs. To avoid path explosion and ensure that constraints are solvable, we only unrolled loops once and ignored loop-continuation condition. The constraints are constructed, processed, and solved based on Z3.

\noindent \textbf{Preprocessing.}
For efficiency, \tool leverages LLVM to compile single operator code file into bitcode and link it with necessary libraries.  

\noindent \textbf{Input Validation Path Extraction.}
We traverse from each operator's entry to explore all paths in validation code. 
If the code of operator is divided into several  blocks with different entries, we extract paths from different code blocks separately and combine them for the input generation process.




\noindent \textbf{Testing.} 
The generated python scripts were executed with target DL library in different modes. We tested TensorFlow's operators in three modes, i.e., native mode, OneDNN mode (executing operators with Intel-optimized operators) and CUDA mode (executing operators with CUDA operators). PyTorch's operators were tested in native mode and CUDA mode. We reproduced all crashes in order to filter out false positives and identified them based on their return code.

\noindent \textbf{Extending for other DL libraries.}
\tool{} is designed to be easily extended to new DL libraries. To support a new DL library, several interfaces in three parts, namely, preprocessing, constraint extraction, and test case generation, need to be implemented. In the preprocessing part, a script is required to extract the operator list and meta info of the operators. We wrote 300 lines of Python code for both TensorFlow and PyTorch. In the constraint extraction part, some input property handling functions must be manually specified to enable \tool{} to understand their mechanism and determine the propagation rules, such as the $GetDim$ function of tensors. Developers can also add more propagation rules for better performance. We wrote 1000 lines of C++ code for both TensorFlow and PyTorch. Finally, in the test case generation part, Python code is needed for building the actual test cases that invoke the specific DL library APIs. This part requires about 100 lines of Python code. Compared to the size of \tool{}, the amount of manual work required to support a new DL library is minimal.
\section{Evaluation}
\subsection{Experiment Setup}

\subsubsection{Research Questions}
Our evaluation aims to answer the following research questions (RQs).
\begin{itemize}[leftmargin=*]
\setlength{\itemsep}{0pt}
\setlength{\parsep}{0pt}
\setlength{\parskip}{0pt}
\item \textbf{RQ1.} \textbf{(Constraint Extraction.)} How effective is \tool in extracting constraints from DL operator code? (\S\ref{sec:rq1})
\item \textbf{RQ2.} \textbf{(Test Case Generation.)} Can the extracted constraints enable \tool to generate high-quality test cases? (\S\ref{sec:rq2})
\item \textbf{RQ3.} \textbf{(Bug Detection.)} 
How is the bug detection capability of \tool? (\S\ref{sec:rq3})
\end{itemize}


\subsubsection{Dataset Selection.}
To investigate the three research questions, we collected operator-level APIs from two of the most widely used DL libraries, TensorFlow 2.9.0 and Pytorch 1.13.0, and applied several exclusion criteria to filter out APIs that \ding{182} have unsupported input parameters such as \code{handler}, \ding{183} cannot be executed independently as they rely on other APIs, \ding{184} are deprecated, \ding{185} only support TPU mode, or \ding{186} are not explicitly bound to an operator. Furthermore, we selected only one of the APIs that called the same operator to prevent redundancy. Following these guidelines, we identified a total of 610 and 485 operator-level APIs for Tensorflow 2.9.0 and PyTorch 1.13.0, respectively.


\subsubsection{Baselines.} 
We compared \tool with several state-of-the-art approaches to assess its effectiveness. 
\begin{itemize}[leftmargin=*]
\setlength{\itemsep}{0pt}
\setlength{\parsep}{0pt}
\setlength{\parskip}{0pt}

    \item DocTer~\cite{DocTer}, a documentation-guided fuzzing technique for testing DL API functions.
    \item DocTer*, a variant of DocTer that we implemented, which transforms DocTer's constraints into ones following the format of \tool and then generates input using the input generation component of \tool\footnote{The input generation component of \tool requires a specific type for each parameter, while DocTer dynamically infers the type during the input generation phase. Therefore, DocTer* uses the types extracted by \tool and obtains other constraints from DocTer's results.}.
    
    \item \toolx, a baseline that we implemented, which leverages the meta information (mentioned in Table~\ref{table:meta-info}) to generates test cases without relying on constraints extracted from the source code.

    \item FreeFuzz~\cite{Wei2022FreeLF}, a mutation-based fuzzing technique that extracts DL high-level API usages from open source.
    
    \item DeepREL~\cite{DeepREL}, a fuzzing technique that infers relational high-level APIs by analyzing APIs sharing similar input and outputs.
\end{itemize}

To answer RQ1, we compared the constraints extracted by \tool and DocTer, as other tools were unable to extract constraints, and we manually analyzed the recall and precision of \tool's constraint extraction.
To answer RQ2, we compared the branch coverage and pass rate of \tool with DocTer, DocTer* and \toolx, as FreeFuzz and DeepREL concentrate on high-level APIs and cannot be easily adapted to our test targets.
To answer RQ3, we compare the bugs detected by \tool with the other four tools.

\subsubsection{Metrics.} 
 To assess \tool's and DocTer's constraint extraction capabilities, we evaluated them with the following metric.
\begin{itemize}[leftmargin=*]
\setlength{\itemsep}{0pt}
\setlength{\parsep}{0pt}
\setlength{\parskip}{0pt}
    \item \textbf{Constraint Number of the Property} (\textbf{CNP}): 
    The CNP indicates the  frequency of a specific property being constrained among all extracted constraints. 
    The CNP for property $dtype$ of tensor is calculated using the following equation:
    \begin{equation} CNP = \sum_{p \in tensors} IsConstrained(p.dtype) \end{equation}
    Here, $tensors$ is a set of all parameters with tensor type among all the constraints.
    By having more properties being constrained and each property being constrained more frequently, we can indicate a higher diversity of the constraints present.

\end{itemize}
To evaluate the quality of test cases, we execute the generated test cases with target DL library, and obtain the quality in two aspects:
\begin{itemize}[leftmargin=*]
\setlength{\itemsep}{0pt}
\setlength{\parsep}{0pt}
\setlength{\parskip}{0pt}
    \item \textbf{Pass Rate}: Pass rate evaluates ratio of generated test cases that can pass all input validation checks. And therefore demonstrates the validity of generated test cases~\cite{DocTer}. Pass rate of a test round for an API can be calculated with the following equation:
        \begin{equation} PassRate =  \frac{api.validate\_cases.size()}{api.test\_cases.size()} \end{equation}
    \item \textbf{Branch Coverage}: Branch coverage is widely adopted in software testing for measuring the quality of the test cases~\cite{Chen2019MatryoshkaFD}. It reveals the path exploration ability of test cases.
\end{itemize}

\subsubsection{Evaluation Configuration.} Our experiments were run on a machine with Intel Xeon CPU (24 cores, 2.20 GHz), NVIDIA V100 GPU, Ubuntu 20.04, and Python 3.8. We test 2000 rounds for each API until (1) the test time of the API reached 30 minutes; (2) 50 crashes or round timeouts are triggered. As DocTer generates VI (Violating Input) and CI (Conforming Input) each for 1000 rounds, we compute its pass rate only for CIs and branch coverage for both VIs and CIs.
In RQ2 and RQ3, all experiments are conducted 5 times to reduce the impact of randomness. For RQ2, the reported results are the average of 5 test runs. For RQ3, we used all bugs detected in 5 test runs as the final results. In addition, to ensure fair comparison, FreeFuzz and DeepREL, which target high-level APIs, were configured to run 2000 times for each API and were limited to run within 72 hours (all other tools can complete the test runs within 72 hours).

\subsection{RQ1: Effectiveness of Constraint Extraction}\label{sec:rq1}

\subsubsection{Comparison with DocTer.}
DocTer utilizes documents to extract API constraints. And DocTer was only able to extract constraints for 608 APIs in tensorflow and 205 APIs in PyTorch due to the unavailable of documents for other APIs. To ensure a fair comparison, we only considered the constraints extracted from these APIs.

Table~\ref{table:CNP} presents the CNP results. 
\tool extracted 3280 constraints in total, \textit{almost} twice the number of constraints extracted by DocTer, which is 1670.
In terms of parameter properties, \tool is more effective than DocTer, especially in terms of the \textit{length} and \textit{value} properties of the \textit{list} type, where DocTer fails to extract any constraints.
Since the \textit{tensor} is the most common parameter of the operator, it has the largest number of corresponding constraints. Moreover, $dtype$-related constraints can be expressed in simple natural language, so DocTer can easily obtain such constraints from the documents, which leads to the similar constraint numbers of \textit{dtype} in the table.



\begin{table}[!ht]
\caption{CNP of constraints extracted by DocTer and \tool}
    \centering
    \label{table:CNP}
    \vspace{-8pt}
    \renewcommand\arraystretch{1.35}
    \begin{adjustbox}{max width=0.48\textwidth}
    \begin{tabular}{c|c|c|c|c|c|c|c|c}
    \hline
        \textbf{Parameter Type} & \textbf{Base Type} & \multicolumn{3}{c|}{\textbf{list}}   & \multicolumn{3}{c|}{\textbf{tensor}} & \multirow{2}{*}{\textbf{Total}} \\ \cline{1-8}
        \textbf{Property} & \textbf{value} & \textbf{dtype} & \textbf{length} & \textbf{value} & \textbf{dtype} & \textbf{dim} & \textbf{shape} & \\ \hline
        \textbf{DocTer} & 220 & 256 & 0 & 0 & 919 & 212 & 63 & 1670 \\ \hline
        \textbf{\tool} & 402 & 283 & 176 & 68 & 1168 & 639 & 554 & 3280\\ \hline
    \end{tabular}
    \end{adjustbox}
\end{table}


\subsubsection{Manual Validation.}
To examine the completeness and accuracy of the extracted constraints by \tool, we randomly selected 25 operators from each of TensorFlow and PyTorch, and conducted a manual verification.
The manual analysis was conducted independently by three authors of this paper, and the final results reached unanimous agreement among all three authors.
The constraints for each operator consist of multiple SMT scripts, with each script corresponding to a path in the input validation code of the operator that passes the check. As some operators have hundreds of scripts, each contains hundreds of lines of code.
For example, \textit{Conv2DOp} of TensorFlow has 104 scripts, with 998 lines of code each\footnote{The constraints of \textit{Conv2DOp} extracted by \tool can be downloaded at \url{https://github.com/shijy16/ACETest}}.
Manually analyzing all of them would be too labor-intensive.
Hence, in cases where the number of scripts for an operator exceeded 10, we randomly chose 10 scripts for verification purposes (64\% of the operators had 10 or fewer scripts).
We evaluate the extracted constraints in two dimensions: \ding{182} The integrity of individual constraint scripts ($Recall_s$), calculated by determining the proportion of satisfied checks within the script's corresponding path. \ding{183} The coverage of all constraints on the checking paths in an operator ($Recall_{op}$). Precision in the extracted constraints and the time cost for verifying each operator are also recorded.
 \vspace{2mm}
\begin{table}[tb]
\caption{Manual analysis result for constraints extraction}
    \centering
    \label{table:cons_check}
    \vspace{-8pt}
    \begin{adjustbox}{max width=0.45\textwidth}
    \begin{tabular}{c|c|c|c|c}
    \hline
        &\bm{$Recall_{s}$}&\bm{$Recall_{op}$}&\textbf{Precision} & \textbf{Avg. Cost(min)}
         \\
    \hline
         \textbf{TensorFlow} & 87\% & 90\% & 100\% &  24 \\ 
    \hline
          \textbf{PyTorch} & 68\% & 75\% & 100\%  & 23 \\  \hline
    \end{tabular}
    \end{adjustbox}
\end{table}

The results are shown in \tabl ~\ref{table:cons_check}. Overall, \tool is able to extract complete constraints for most operators in TensorFlow and PyTorch and cover all paths with 100\% precision.
However, constraints in some operators are not fully extracted. 
We examined the operators with incomplete constraint extraction and identified several main reasons:
\ding{182} Loops in the input validation code: We only unrolled the loop once and ignored the loop condition to avoid false positives, which also resulted in missing constraints within the loop.
\ding{183} Incomplete compilation: Due to the large size of DL libraries, we could not perform static analysis on the entire code base. As a fallback, we only compiled the code file of each single operator and linked it with necessary code files, resulting in some functions not being linked.
\ding{184} C++ pointers: Despite the implementation of virtual memory in \tool to handle pointers, the inaccuracy of current static analysis can still result in false negatives. Additionally, \tool is currently unable to retrieve function pointers. 
\ding{185} Constraints on \textit{tensor}'s \textit{value} or \textit{number of elements}: These properties are out of the scope of \tool's constraint model, as they are rare and too expensive too handle.
Moreover, the manual verification process of constraints for each operator required an average of 23-24 minutes, and manually extracting constraints could take up to 10 times longer. The manual extraction process involves identifying each path, verifying the logic in each path, and formatting the constraints into SMT scripts. In contrast, \tool can extract constraints for most operators in 1 minute.



\vspace{-5pt}
\begin{tcolorbox}[colback=blue!5!white,enhanced,frame hidden, boxsep=0pt,left=5pt,right=5pt,top=2pt,bottom=2pt]
\textbf{Result for RQ1:} 
Compared to DocTer,
\tool is capable of extracting more constraints, and its precision and recall in constraints extraction are satisfactory.
\end{tcolorbox}
\vspace{-5pt}
\subsection{RQ2: Effectiveness of Test Case Generation}\label{sec:rq2}

To investigate the effectiveness of test case generation, we compared \tool against three baselines (\textit{i.e.}, \textit{\toolx}, \textit{DocTer}, and \textit{DocTer*}) in terms of  pass rate and branch coverage. 

\vspace{-5pt}
\subsubsection{Pass Rate.}

Table~\ref{table:pass_rate} shows the pass rate results of different techniques.
We observe that \tool significantly outperforms DocTer, DocTer* and \toolx, achieving the state-of-the-art result of 32.51\% and 43.66\% pass rate on TensorFlow and PyTorch.
Compared with DocTer, \tool increased the pass rate by 162.60\% and 138.58\% on TensorFlow and PyTorch, respectively.
It also achiev-es significant pass rate improvements compared to DocTer* (47.73\% and 39.22\% on TensorFlow and PyTorch) and outperforms \toolx by 34.51\% and 36.65\%.
These results demonstrate \tool's ability to generate valid test cases that thoroughly explore the core functional code.

\begin{table}[h]
\caption{Comparison on pass rate}
    \centering
    \label{table:pass_rate}
    \renewcommand\arraystretch{1.35}
    \vspace{-8pt}
    \begin{adjustbox}{max width=0.45\textwidth}
    \begin{tabular}{c|c|c|c|c}
    \hline
         &\textbf{DocTer*}& \textbf{DocTer} & \textbf{\toolx}&\textbf{\tool}
         \\
    \hline
         \textbf{TensorFlow} & 22.00\% & 12.38\%  & 24.17\% & \textbf{32.51\%}   \\ 
    \hline
          \textbf{PyTorch} & 31.36\% & 18.3\% & 31.95\% & \textbf{43.66\%} \\  
    \hline
    \end{tabular}
    \end{adjustbox}
    \vspace{-10pt}
\end{table}


\subsubsection{Branch Coverage.}
\begin{figure}[tb]
     \setlength{\abovecaptionskip}{1pt}
    \centering
    \includegraphics[width=0.4\textwidth]{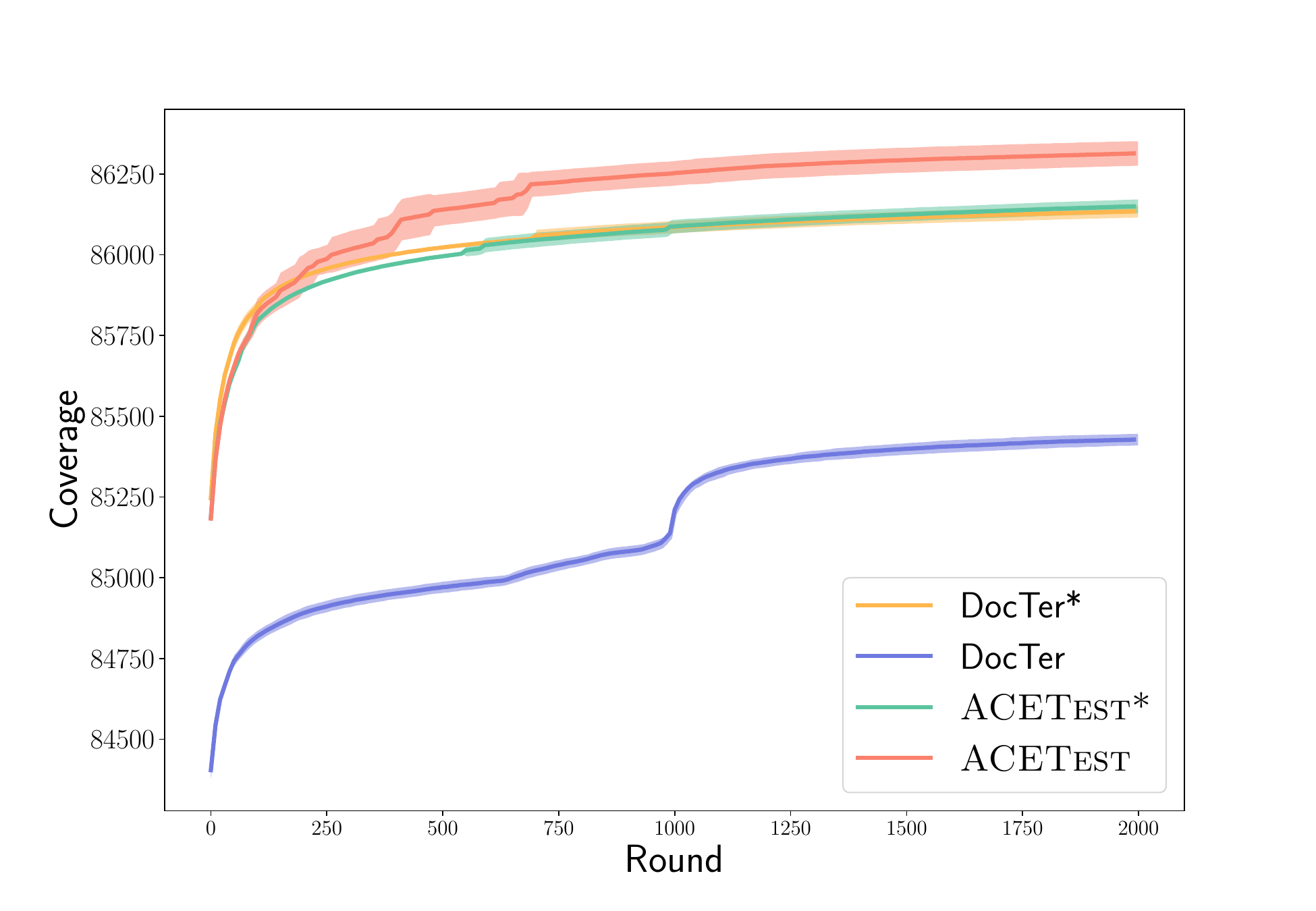}
     \caption{The average branch coverage for each operator compared with baselines. The curve represents the average of 5 tests, with a 95\% confidence interval.}
    \label{fig:cov}
    \vspace{-3pt}
\end{figure}

\figu ~\ref{fig:cov} presents the average branch coverage of each operator. 
It demonstrates that among the four tools, \tool achieves the highest branch coverage. And
the coverage of \tool keeps growing as the number of testing rounds increases, and the upward trend persists until the end of the testing process.
The results suggest that \tool has a higher branch coverage in DL operators and continuously generates diverse test cases that explore new paths.


\tabl ~\ref{table:pass_rate} and \figu ~\ref{fig:cov} indicate that DocTer had relatively lower overall pass rate and branch coverage. This can be attributed to the incorrect and incomplete extraction of parameter types from the documents, resulting in lower coverage and pass rate of certain operators. In contrast, the other three tools use the parameter types obtained by \tool from the operator registration, which are less prone to errors. Additionally, at the 1000th round, DocTer switches from generating test cases that satisfy the constraints to generating test cases that violate the constraints, which leads to a sharp increase in coverage.

\vspace{-5pt}
\begin{tcolorbox}[colback=blue!5!white,enhanced,frame hidden, boxsep=0pt,left=5pt,right=5pt,top=2pt,bottom=2pt]
\textbf{Result for RQ2:} 
Compared to the three baselines (\textit{i.e.}, DocTer, DocTer*, and \toolx), \tool generates a more varied and valid set of test cases with the aid of extracted constraints.
\end{tcolorbox}

\vspace{-15pt}
\subsection{RQ3: Effectiveness of Bug Detection}\label{sec:rq3}

\begin{figure}[tb]
    \setlength{\abovecaptionskip}{1pt}
    \centering
    \includegraphics[width=0.45\textwidth]{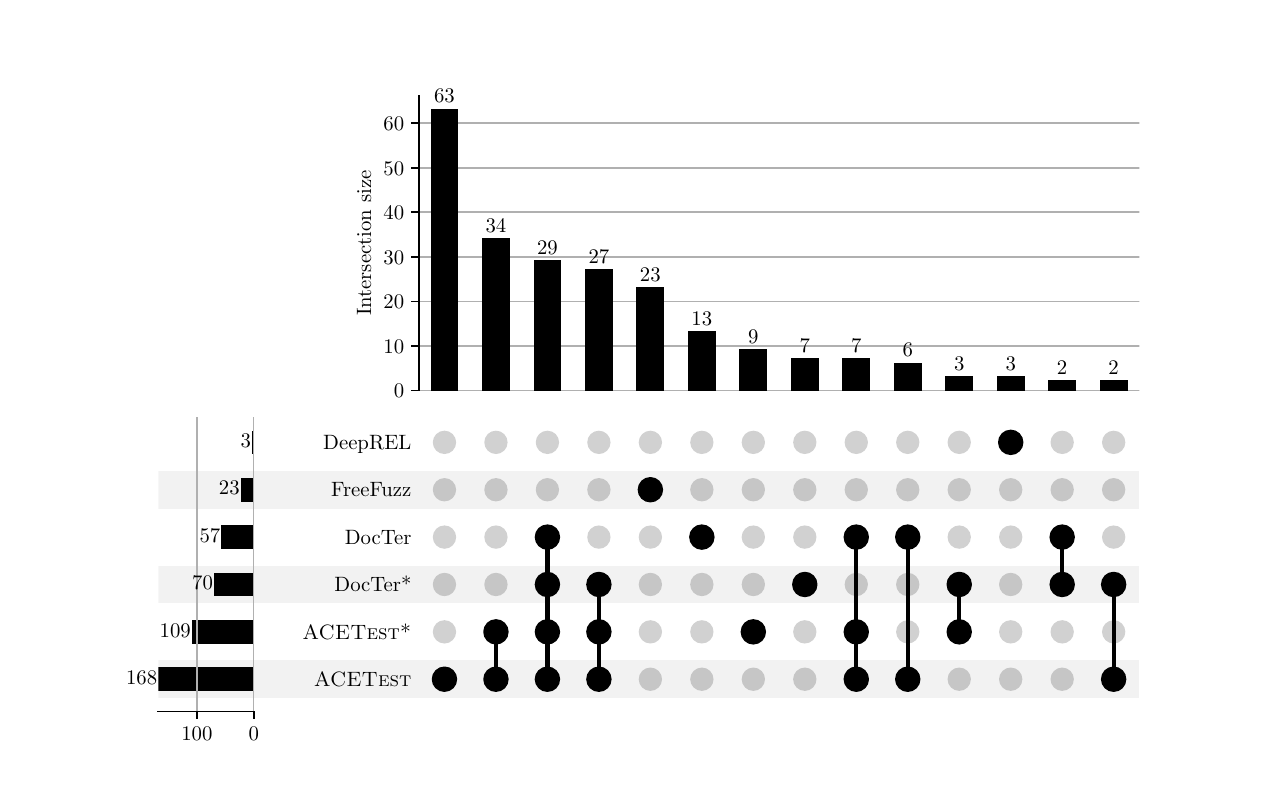}
     \caption{The Effectiveness of Bug Detection}
    \label{fig:bugs}
\end{figure}

In this section, we performed an evaluation to assess the effectiveness of bug detection.
The overall evaluation results are listed in \figu~\ref{fig:bugs} and \tabl~\ref{table:bugs}. 

\begin{table}[tb]
\setlength{\abovecaptionskip}{1pt}
  \caption{The Distribution of Different Types of Bugs Detected in TensorFlow and PyTorch}
    \label{table:bugs}
    \centering
    \renewcommand\arraystretch{1.35}
    \begin{adjustbox}{max width=0.46\textwidth}
    \begin{tabular}{c|cccc|cccc|c}
\hline
\multicolumn{1}{c|}{\multirow{2}{*}{\textbf{Tools}}} & \multicolumn{4}{c|}{\textbf{TensorFlow}}                                                                                              & \multicolumn{4}{c|}{\textbf{PyTorch}}                                                                                                 & \multicolumn{1}{c}{\multirow{2}{*}{\textbf{Total}}} \\ \cline{2-9}
\multicolumn{1}{l|}{}                           & \multicolumn{1}{l|}{\textbf{ABRT}} & \multicolumn{1}{l|}{\textbf{FPE}} & \multicolumn{1}{l|}{\textbf{SEGV}} & \textbf{Total} & \multicolumn{1}{c|}{\textbf{ABRT}} & \multicolumn{1}{c|}{\textbf{FPE}} & \multicolumn{1}{c|}{\textbf{SEGV}} & \textbf{Total} & \multicolumn{1}{c}{}                                \\ \hline
\textbf{FreeFuzz}                                & \multicolumn{1}{c|}{22}               & \multicolumn{1}{c|}{0}               & \multicolumn{1}{c|}{0}                & 22             & \multicolumn{1}{c|}{0}                & \multicolumn{1}{c|}{0}               & \multicolumn{1}{c|}{1}                & 1              & 23                                                   \\ \hline
\textbf{DeepREL}                                 & \multicolumn{1}{c|}{2}                & \multicolumn{1}{c|}{1}               & \multicolumn{1}{c|}{0}                & 3              & \multicolumn{1}{c|}{0}                & \multicolumn{1}{c|}{0}               & \multicolumn{1}{c|}{0}                & 0              & 3                                                    \\ \hline
\textbf{DocTer*}                                 & \multicolumn{1}{c|}{42}               & \multicolumn{1}{c|}{2}               & \multicolumn{1}{c|}{21}               & 63             & \multicolumn{1}{c|}{0}                & \multicolumn{1}{c|}{0}               & \multicolumn{1}{c|}{5}                & 5              & 68                                                   \\ \hline
\textbf{DocTer}                                  & \multicolumn{1}{c|}{42}               & \multicolumn{1}{c|}{0}               & \multicolumn{1}{c|}{15}               & 57             & \multicolumn{1}{c|}{0}                & \multicolumn{1}{c|}{0}               & \multicolumn{1}{c|}{0}                & 0              & 57                                                   \\ \hline
\textbf{\toolx}                                     & \multicolumn{1}{c|}{54}               & \multicolumn{1}{c|}{0}               & \multicolumn{1}{c|}{24}               & 78             & \multicolumn{1}{c|}{1}                & \multicolumn{1}{c|}{1}               & \multicolumn{1}{c|}{29}               & 31             & 109                                                  \\ \hline
\textbf{\tool}                    & \multicolumn{1}{c|}{\textbf{85}}      & \multicolumn{1}{c|}{\textbf{11}}     & \multicolumn{1}{c|}{\textbf{35}}      & \textbf{131}   & \multicolumn{1}{c|}{\textbf{2}}       & \multicolumn{1}{c|}{\textbf{2}}      & \multicolumn{1}{c|}{\textbf{33}}      & \textbf{37}    & \textbf{168}                                         \\ \hline
\end{tabular}
    \end{adjustbox}
    \vspace{5pt}
\end{table}

\vspace{-6pt}
\subsubsection{Result Analysis.}
In particular, as we can see from Figure~\ref{fig:bugs}, the bar chart in the left-hand side illustrates the total number of bugs found by each tool. The two vertical bar charts together present  the number of bugs (see the upper vertical bar chart) that can be detected uniquely by the corresponding tool(s) (see the lower vertical bar chart with lined black dots to represent the use of the corresponding tool(s)). \tool significantly outperformed \textit{ALL} 
five state-of-the-art tools. It  could detect the most 168 bugs and further detected 63 unique bugs, which is a 37.50\% improvement.  \toolx performed the second, by finding 109 ($47.81\%$) while 
the four remaining tools, namely DeepREL, FreeFuzz, DocTer, and DocTer*, were less 
effective, reporting between 3 and 70 bugs, with an average of 38.25 ($\frac{3+23+57+70}{4}=38.25$)  bugs.


Besides, we analyzed the distribution of three types of crashes, namely \textit{Abort} (ABRT), \textit{Floating Point Exception} (FPE),  and \textit{Segmentation Fault} (SEGV), over all bugs. The results are presented in \tabl ~\ref{table:bugs}. The findings indicate that \tool detected more bugs of each crash type in both TensorFlow and PyTorch compared to other tools. 
Specifically, \tool was able to detect 1.95 to 55 times more bugs than SOTA techniques (e.g., DeepREL and DocTer). 
Among all the bugs detected by \tool, 29 bugs in PyTorch and 79 bugs in TensorFlow were new, of which 87 were confirmed by the developers, and 35 were fixed.

It is noteworthy that both DeepREL and FreeFuzz exhibit relatively lower performance compared to their original results. After careful investigation, two primary factors were identified:
\ding{182} Test oracle. Both approaches employ distinct test oracles to identify bugs, excluding crash bugs. FreeFuzz detects wrong-computation bugs and performance bugs, while DeepREL additionally targets relational API bugs. To ensure a fair comparison, our experiments only consider crash bugs for all tools.
\ding{183} Target version. The versions of TensorFlow and PyTorch used in their respective papers are significantly older than the version we utilized for testing. Consequently, the update in version has led to a decrease in the occurrence of superficial bugs.

\vspace{-6pt}
\subsubsection{Case Study.} 
\begin{figure}[tb]
\setlength{\abovecaptionskip}{1pt}
    \centering
    \includegraphics[width=\linewidth]{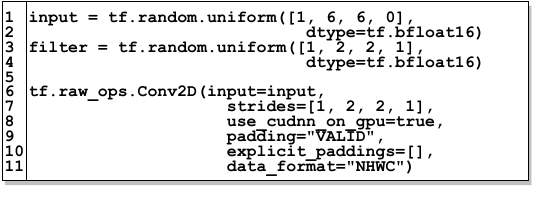}
    \caption{PoC for Conv2D in TensorFlow}
    \label{fig:case}
    \vspace{-3pt}
\end{figure}

Furthermore, we conducted a case study to illustrate the efficacy of \tool. 
The bug resided in the operator \code{Conv2D} of TensorFLow. The buggy input is shown in \figu ~\ref{fig:case}. \code{Conv2D} consists of 7 parameters and complex input constraints. For instance, when \code{data_format} is "NHWC", the first and third elements of \code{strides} must be 1. Furthermore, the shape of output tensor is calculated and validated using intricate logic, with at least 5 parameters involved in the calculation. None of the other tools were able to generate valid test cases that passed the checks and entered the functional code, where the bug was located. Specifically, the bug could be triggered when the functional code was executed with \code{input} having zero elements. Only \tool was able to generate valid test cases and trigger the bug. This case study demonstrates \tool's superiority in extracting constraints, as this bug is detected only by \tool.

\vspace{-6pt}
\subsubsection{Failure Analysis.} Although \tool detected 63 unique bugs that were not found by the five baselines, it missed 60 bugs that these tools were able to detect. After analyzing these bugs, we identified four main reasons for \tool's failure to detect them:
\ding{182} FreeFuzz and DeepREL have different test targets than ACETest, resulting in unique bugs that they are able to detect.
\ding{183} DocTer implements an input mutation technique that can generate more edge cases, allowing it to find bugs that other tools may miss. 
\ding{184} Since all tools generate test cases with randomness, the results can be unpredictable.
\ding{185} Although input validation code has fewer bugs than functional code, some bugs may still be present. However, \tool only generates test cases to test the core functional code, not input validation code, which could lead to missing bugs.

\vspace{-5pt}
\begin{tcolorbox}[colback=blue!5!white,enhanced,frame hidden, boxsep=0pt,left=5pt,right=5pt,top=2pt,bottom=2pt]
\textbf{Result for RQ3:} 
\tool  outperforms state-of-the-art approaches in terms of bug detection. In particular, \tool found the most 168  ($73.68\%$) bugs and further found 63 unique ones, which is a 37.50\% improvement.
\end{tcolorbox}

\vspace{-8pt}
\section{DISCUSSION}
\noindent{\textbf{Threats to Validity.}} A potential threat to the validity of our study is the randomness of the testing process, which may lead to our results being specific to certain cases. To mitigate this risk, we conducted five rounds of testing and generated 2000 test cases for each API in each round to ensure that our results are consistent and not influenced by randomness. However, it is still possible that our testing results may not fully capture the behavior of the tools under all possible scenarios. Another potential threat to the validity of our study is the subjectivity of the manual analysis process. To address this risk, we had three experts conduct independent analyses, and we cross-validated the results to reduce the impact of subjectivity on our findings. Finally, \tool relies on the SVF tool to extract the VFG and CVG, which may have implementation flaws that could affect the accuracy of our results.


\noindent{\textbf{Bug Identification.}} When performing DL operator testing, we only consider crashes as bugs and ignore some bugs that do not crash, such as bugs with incorrect functional implementation. These bugs may be detected with differential testing used in~\cite{Muffin, Wei2022FreeLF}. Besides, with ASAN~\cite{Jeon2020FuZZanES} enabled, we may detect more bugs.


\vspace{-10pt}
\section{RELATED WORK}


\textbf{DL Operators Testing.}
DL operators are the fundamental components of DL libraries, bugs in operators may affect the behavior of DL models. As a consequence, several approaches have been proposed~\cite{Duo, Predoo, Wei2022FreeLF}.
To detect precision bugs in operators, Predoo~\cite{Predoo} treats DL operators testing problem as a searching problem to maximize output precision errors. 
FreeFuzz~\cite{Wei2022FreeLF} extracts API usage via mining from open source. It collects the types and values of each parameter, and shapes of input/output tensors by running all the code/models, and performs differential testing for each API.
DocTer~\cite{DocTer} proposes to extract DL-specific input constraints for DL API functions by analyzing API documentation. With a set of manually annotated API documents with precise constraint information, DocTer can derive constraint extraction rules, which can be used to predict API parameter constraints.
DeepREL~\cite{DeepREL} focuses on inferring relational APIs by analyzing APIs sharing similar input parameters and outputs. With formalizing the notion of value equivalence and status equivalence for relational APIs, it can find potential inconsistencies with testing.
IvySyn~\cite{ivysyn} utilizes the static typing feature of operators' native code (C/C++) through instrumentation and direct of operators' native APIs. As a result, it can generate inputs with the correct type, which sets it apart from other Python API testing approaches.

These methods share the same goal with ours. They extract API usage or operator constraints from open source or document, which are limited or incorrect. Instead, we extract constraints by analyzing code of DL libraries. These constraints are more precise, and can express the relations among each parameter.

\noindent \textbf{DL Library Testing.} 
To detect bugs in DL libraries, multiple techniques~\cite{Nejadgholi2019ASO, Guo2020AudeeAT, Wang2020DeepLL, Luo2021GraphBasedFT, Wang2022EAGLECE, Muffin, Wu2022DeepCovCG,Yang2023FuzzingAD} have been proposed to test them by generating or mutating DL models. 
Luo et al.~\cite{Luo2021GraphBasedFT} design a graph-based fuzz testing method, which introduced six different mutations by exploring combinations of model structures, parameters, and data inputs.
EAGLE~\cite{Wang2022EAGLECE} proposes to test a single DL library with generating equivalent graphs, which use different APIs, data types, or optimizations to achieve the same functionality.
CRADLE~\cite{Pham2019CRADLECV} takes pre-trained model as input and performs cross-implementation inconsistency checking to detect bugs.
Instead of focusing on detecting bugs in model inference phase, Muffin~\cite{Muffin} proposes to make differential testing feasible in the model training phase by tailoring a set of metrics to measure the inconsistencies between different DL libraries.
MEMO~\cite{Li2022MEMOCM} proposes a coverage based model generation technique with a customized Markov Chain Monte Carlo algorithm to explore new layer types, layer pairs, and layer parameters.
TitanFuzz~\cite{Deng2022FuzzingDL} directly adopts large pre-trained Language model to generate input programs for fuzzing DL libraries.

These approaches focus on detecting either logic bugs by differential testing in single or multiple DL libraries, or memory bugs by monitoring crashes. One common challenge is how to generate various models. Instead, we pay attention to detect bugs in operators, and try to generate diverse parameter inputs of operators, which are different from model.

\noindent \textbf{Symbolic Execution.}
To test program deeper, many researchers proposed to augment fuzzing through symbolic execution~\cite{Stephens2016DrillerAF,Noller2018BadgerCA,Wu2018FUZETF,Peng20191dVulD1}. However, symbolic execution suffers from path explosion problem and constraint solving limitations. To make it more practical, David et al.~\cite{Trabish2018ChoppedSE} represents chopped symbolic execution that leverages various on-demand static analyses at runtime to automatically exclude code fragments. 
Specifically, it allows users to specify uninteresting parts of code to exclude during the analysis. 
PHchop~\cite{Singh2020ParallelCS} introduces a parallel approach to chopped symbolic execution.
Learch~\cite{He2021LearningTE} proposes a learning-based strategy, that can select promising states for symbolic execution. Cha et al.~\cite{Cha2022EnhancingDS} proposes a method that can automatically learn search heuristics by defining a class of search heuristics and finding an optimal heuristic for each subject program efficiently. 
\textit{Directed Symbolic Execution} (DSE) is a technique that explores specific paths and collects and solves the relevant constraints instead of all possible paths. It is used in various testing techniques, such as DART ~\cite{DART}, to reduce the limitations of symbolic execution. DSE is similar to the constraint extraction technique of \tool{}.
However, while DSE extracts constraints for all values, \tool focuses on constraints for certain types of values.

\vspace{-8pt}
\section{Conclusion}
In this paper, we present a technique called \tool{}, which can automatically extract input validation constraints to aid DL operator testing.
Specifically, \tool{} can automatically identify the input validation code in DL operators, extract the related constraints and build test cases according to the constraints.
\tool{} has demonstrated superior performance on popular DL libraries, TensorFlow and PyTorch.
Experiments results demonstrate that \tool is capable of extracting 96.4\% more constraints than DocTer, which is a SOTA technique for extracting constraints from documentation.
Moreover, \tool can detect 1.95 to 55 times more bugs than SOTA techniques.
In total, we have used \tool to detect 108 previously unknown bugs on TensorFlow and PyTorch, with 87 of them confirmed and five of them assigned with CVE IDs.

\vspace{-8pt}
\section*{Acknowledgment}
The authors would like to thank the anonymous reviewers for their helpful feedback on an earlier version of this paper. 
This work is partly supported by National Key R\&D Program of China under Grant \#2022YFB3103900, Strategic Priority Research Program of the CAS under Grant \#XDCO2030200 and Chinese National Natural Science Foundation (Grants \#62032010, \#62202462).


\bibliographystyle{ACM-Reference-Format}
\bibliography{ref}


\begin{thebibliography}{54}


\ifx \showCODEN    \undefined \def \showCODEN     #1{\unskip}     \fi
\ifx \showDOI      \undefined \def \showDOI       #1{#1}\fi
\ifx \showISBNx    \undefined \def \showISBNx     #1{\unskip}     \fi
\ifx \showISBNxiii \undefined \def \showISBNxiii  #1{\unskip}     \fi
\ifx \showISSN     \undefined \def \showISSN      #1{\unskip}     \fi
\ifx \showLCCN     \undefined \def \showLCCN      #1{\unskip}     \fi
\ifx \shownote     \undefined \def \shownote      #1{#1}          \fi
\ifx \showarticletitle \undefined \def \showarticletitle #1{#1}   \fi
\ifx \showURL      \undefined \def \showURL       {\relax}        \fi
\providecommand\bibfield[2]{#2}
\providecommand\bibinfo[2]{#2}
\providecommand\natexlab[1]{#1}
\providecommand\showeprint[2][]{arXiv:#2}

\bibitem[Abadi et~al\mbox{.}(2016)]%
        {TensorFlow}
\bibfield{author}{\bibinfo{person}{Mart{\'i}n Abadi}, \bibinfo{person}{Paul
  Barham}, \bibinfo{person}{Jianmin Chen}, \bibinfo{person}{Z. Chen},
  \bibinfo{person}{Andy Davis}, \bibinfo{person}{Jeffrey Dean},
  \bibinfo{person}{Matthieu Devin}, \bibinfo{person}{Sanjay Ghemawat},
  \bibinfo{person}{Geoffrey Irving}, \bibinfo{person}{Michael Isard},
  \bibinfo{person}{Manjunath Kudlur}, \bibinfo{person}{Josh Levenberg},
  \bibinfo{person}{Rajat Monga}, \bibinfo{person}{Sherry Moore},
  \bibinfo{person}{Derek~Gordon Murray}, \bibinfo{person}{Benoit Steiner},
  \bibinfo{person}{Paul~A. Tucker}, \bibinfo{person}{Vijay Vasudevan},
  \bibinfo{person}{Pete Warden}, \bibinfo{person}{Martin Wicke},
  \bibinfo{person}{Yuan Yu}, {and} \bibinfo{person}{Xiaoqiang Zhang}.}
  \bibinfo{year}{2016}\natexlab{}.
\newblock \showarticletitle{TensorFlow: A system for large-scale machine
  learning}.
\newblock \bibinfo{journal}{\emph{12th {USENIX} Symposium on Operating Systems
  Design and Implementation, {OSDI}}} (\bibinfo{year}{2016}).
\newblock


\bibitem[Ahmad et~al\mbox{.}(2020)]%
        {Ahmad2020ATA}
\bibfield{author}{\bibinfo{person}{Wasi~Uddin Ahmad}, \bibinfo{person}{Saikat
  Chakraborty}, \bibinfo{person}{Baishakhi Ray}, {and} \bibinfo{person}{Kai-Wei
  Chang}.} \bibinfo{year}{2020}\natexlab{}.
\newblock \showarticletitle{A Transformer-based Approach for Source Code
  Summarization}.
\newblock \bibinfo{journal}{\emph{ArXiv}}  \bibinfo{volume}{abs/2005.00653}
  (\bibinfo{year}{2020}).
\newblock


\bibitem[Cadar et~al\mbox{.}(2011)]%
        {Cadar2011SymbolicEF}
\bibfield{author}{\bibinfo{person}{Cristian Cadar}, \bibinfo{person}{Patrice
  Godefroid}, \bibinfo{person}{Sarfraz Khurshid}, \bibinfo{person}{Corina~S.
  Pasareanu}, \bibinfo{person}{Koushik Sen}, \bibinfo{person}{Nikolai
  Tillmann}, {and} \bibinfo{person}{Willem Visser}.}
  \bibinfo{year}{2011}\natexlab{}.
\newblock \showarticletitle{Symbolic execution for software testing in
  practice: preliminary assessment}.
\newblock \bibinfo{journal}{\emph{2011 33rd International Conference on
  Software Engineering (ICSE)}} (\bibinfo{year}{2011}),
  \bibinfo{pages}{1066--1071}.
\newblock


\bibitem[Cai et~al\mbox{.}(2022)]%
        {Cai2022PeahenFA}
\bibfield{author}{\bibinfo{person}{Yuandao Cai}, \bibinfo{person}{Chengfeng
  Ye}, \bibinfo{person}{Qingkai Shi}, {and} \bibinfo{person}{Charles Zhang}.}
  \bibinfo{year}{2022}\natexlab{}.
\newblock \showarticletitle{Peahen: fast and precise static deadlock detection
  via context reduction}.
\newblock \bibinfo{journal}{\emph{Proceedings of the 30th ACM Joint European
  Software Engineering Conference and Symposium on the Foundations of Software
  Engineering}} (\bibinfo{year}{2022}).
\newblock


\bibitem[Cha et~al\mbox{.}(2022)]%
        {Cha2022EnhancingDS}
\bibfield{author}{\bibinfo{person}{Sooyoung Cha}, \bibinfo{person}{Seongjoon
  Hong}, \bibinfo{person}{Jiseong Bak}, \bibinfo{person}{Jingyoung Kim},
  \bibinfo{person}{Junhee Lee}, {and} \bibinfo{person}{Hakjoo Oh}.}
  \bibinfo{year}{2022}\natexlab{}.
\newblock \showarticletitle{Enhancing Dynamic Symbolic Execution by
  Automatically Learning Search Heuristics}.
\newblock \bibinfo{journal}{\emph{IEEE Transactions on Software Engineering}}
  \bibinfo{volume}{48} (\bibinfo{year}{2022}), \bibinfo{pages}{3640--3663}.
\newblock


\bibitem[Chen et~al\mbox{.}(2019)]%
        {Chen2019MatryoshkaFD}
\bibfield{author}{\bibinfo{person}{Peng Chen}, \bibinfo{person}{Jianzhong Liu},
  {and} \bibinfo{person}{Hao Chen}.} \bibinfo{year}{2019}\natexlab{}.
\newblock \showarticletitle{Matryoshka: Fuzzing Deeply Nested Branches}.
\newblock \bibinfo{journal}{\emph{Proceedings of the 2019 ACM SIGSAC Conference
  on Computer and Communications Security}} (\bibinfo{year}{2019}).
\newblock


\bibitem[Christou et~al\mbox{.}(2022)]%
        {ivysyn}
\bibfield{author}{\bibinfo{person}{Neophytos Christou}, \bibinfo{person}{Di
  Jin}, \bibinfo{person}{Vaggelis Atlidakis}, \bibinfo{person}{Baishakhi Ray},
  {and} \bibinfo{person}{Vasileios~P. Kemerlis}.}
  \bibinfo{year}{2022}\natexlab{}.
\newblock \showarticletitle{IvySyn: Automated Vulnerability Discovery for Deep
  Learning Frameworks}.
\newblock \bibinfo{journal}{\emph{ArXiv}}  \bibinfo{volume}{abs/2209.14921}
  (\bibinfo{year}{2022}).
\newblock


\bibitem[de~Moura and Bj{\o}rner(2008)]%
        {Z3}
\bibfield{author}{\bibinfo{person}{Leonardo~Mendonça de Moura} {and}
  \bibinfo{person}{Nikolaj~S. Bj{\o}rner}.} \bibinfo{year}{2008}\natexlab{}.
\newblock \showarticletitle{Z3: An Efficient SMT Solver}. In
  \bibinfo{booktitle}{\emph{International Conference on Tools and Algorithms
  for Construction and Analysis of Systems}}.
\newblock


\bibitem[Deng et~al\mbox{.}(2022a)]%
        {Deng2022FuzzingDL}
\bibfield{author}{\bibinfo{person}{Yinlin Deng}, \bibinfo{person}{Chun Xia},
  \bibinfo{person}{Haoran Peng}, \bibinfo{person}{Chenyuan Yang}, {and}
  \bibinfo{person}{Lingming Zhang}.} \bibinfo{year}{2022}\natexlab{a}.
\newblock \showarticletitle{Fuzzing Deep-Learning Libraries via Large Language
  Models}.
\newblock \bibinfo{journal}{\emph{ArXiv}}  \bibinfo{volume}{abs/2212.14834}
  (\bibinfo{year}{2022}).
\newblock


\bibitem[Deng et~al\mbox{.}(2022b)]%
        {DeepREL}
\bibfield{author}{\bibinfo{person}{Yinlin Deng}, \bibinfo{person}{Chenyuan
  Yang}, \bibinfo{person}{Anjiang Wei}, {and} \bibinfo{person}{Lingming
  Zhang}.} \bibinfo{year}{2022}\natexlab{b}.
\newblock \showarticletitle{Fuzzing deep-learning libraries via automated
  relational API inference}.
\newblock \bibinfo{journal}{\emph{Proceedings of the 30th ACM Joint European
  Software Engineering Conference and Symposium on the Foundations of Software
  Engineering}} (\bibinfo{year}{2022}).
\newblock


\bibitem[Devlin et~al\mbox{.}(2019)]%
        {BERT}
\bibfield{author}{\bibinfo{person}{Jacob Devlin}, \bibinfo{person}{Ming-Wei
  Chang}, \bibinfo{person}{Kenton Lee}, {and} \bibinfo{person}{Kristina
  Toutanova}.} \bibinfo{year}{2019}\natexlab{}.
\newblock \showarticletitle{BERT: Pre-training of Deep Bidirectional
  Transformers for Language Understanding}.
\newblock \bibinfo{journal}{\emph{ArXiv}}  \bibinfo{volume}{abs/1810.04805}
  (\bibinfo{year}{2019}).
\newblock


\bibitem[Dudina and Belevantsev(2017)]%
        {dudina2017using}
\bibfield{author}{\bibinfo{person}{IA Dudina} {and} \bibinfo{person}{AA
  Belevantsev}.} \bibinfo{year}{2017}\natexlab{}.
\newblock \showarticletitle{Using static symbolic execution to detect buffer
  overflows}.
\newblock \bibinfo{journal}{\emph{Programming and Computer Software}}
  \bibinfo{volume}{43} (\bibinfo{year}{2017}), \bibinfo{pages}{277--288}.
\newblock


\bibitem[Fan et~al\mbox{.}(2019)]%
        {Fan2019SMOKESP}
\bibfield{author}{\bibinfo{person}{Gang Fan}, \bibinfo{person}{Rongxin Wu},
  \bibinfo{person}{Qingkai Shi}, \bibinfo{person}{Xiao Xiao},
  \bibinfo{person}{Jinguo Zhou}, {and} \bibinfo{person}{Charles Zhang}.}
  \bibinfo{year}{2019}\natexlab{}.
\newblock \showarticletitle{SMOKE: Scalable Path-Sensitive Memory Leak
  Detection for Millions of Lines of Code}.
\newblock \bibinfo{journal}{\emph{2019 IEEE/ACM 41st International Conference
  on Software Engineering (ICSE)}} (\bibinfo{year}{2019}),
  \bibinfo{pages}{72--82}.
\newblock


\bibitem[Gu et~al\mbox{.}(2022)]%
        {Muffin}
\bibfield{author}{\bibinfo{person}{Jiatao Gu}, \bibinfo{person}{Xuchuan Luo},
  \bibinfo{person}{Yangfan Zhou}, {and} \bibinfo{person}{Xin Wang}.}
  \bibinfo{year}{2022}\natexlab{}.
\newblock \showarticletitle{Muffin: Testing Deep Learning Libraries via Neural
  Architecture Fuzzing}.
\newblock \bibinfo{journal}{\emph{2022 IEEE/ACM 44th International Conference
  on Software Engineering (ICSE)}} (\bibinfo{year}{2022}),
  \bibinfo{pages}{1418--1430}.
\newblock


\bibitem[Guo et~al\mbox{.}(2020)]%
        {Guo2020AudeeAT}
\bibfield{author}{\bibinfo{person}{Qianyu Guo}, \bibinfo{person}{Xiaofei Xie},
  \bibinfo{person}{Yi Li}, \bibinfo{person}{Xiaoyu Zhang},
  \bibinfo{person}{Yang Liu}, \bibinfo{person}{Xiaohong Li}, {and}
  \bibinfo{person}{Chao Shen}.} \bibinfo{year}{2020}\natexlab{}.
\newblock \showarticletitle{Audee: Automated Testing for Deep Learning
  Frameworks}.
\newblock \bibinfo{journal}{\emph{2020 35th IEEE/ACM International Conference
  on Automated Software Engineering (ASE)}} (\bibinfo{year}{2020}),
  \bibinfo{pages}{486--498}.
\newblock


\bibitem[He et~al\mbox{.}(2021)]%
        {He2021LearningTE}
\bibfield{author}{\bibinfo{person}{Jingxuan He}, \bibinfo{person}{Gishor
  Sivanrupan}, \bibinfo{person}{Petar Tsankov}, {and}
  \bibinfo{person}{Martin~T. Vechev}.} \bibinfo{year}{2021}\natexlab{}.
\newblock \showarticletitle{Learning to Explore Paths for Symbolic Execution}.
\newblock \bibinfo{journal}{\emph{Proceedings of the 2021 ACM SIGSAC Conference
  on Computer and Communications Security}} (\bibinfo{year}{2021}).
\newblock


\bibitem[He et~al\mbox{.}(2016)]%
        {CV1}
\bibfield{author}{\bibinfo{person}{Kaiming He}, \bibinfo{person}{X. Zhang},
  \bibinfo{person}{Shaoqing Ren}, {and} \bibinfo{person}{Jian Sun}.}
  \bibinfo{year}{2016}\natexlab{}.
\newblock \showarticletitle{Deep Residual Learning for Image Recognition}.
\newblock \bibinfo{journal}{\emph{2016 IEEE Conference on Computer Vision and
  Pattern Recognition (CVPR)}} (\bibinfo{year}{2016}),
  \bibinfo{pages}{770--778}.
\newblock


\bibitem[Jeon et~al\mbox{.}(2020)]%
        {Jeon2020FuZZanES}
\bibfield{author}{\bibinfo{person}{Yuseok Jeon}, \bibinfo{person}{Wookhyun
  Han}, \bibinfo{person}{Nathan Burow}, {and} \bibinfo{person}{Mathias Payer}.}
  \bibinfo{year}{2020}\natexlab{}.
\newblock \showarticletitle{FuZZan: Efficient Sanitizer Metadata Design for
  Fuzzing}. In \bibinfo{booktitle}{\emph{USENIX Annual Technical Conference}}.
\newblock


\bibitem[King(1976)]%
        {king1976symbolic}
\bibfield{author}{\bibinfo{person}{James~C King}.}
  \bibinfo{year}{1976}\natexlab{}.
\newblock \showarticletitle{Symbolic execution and program testing}.
\newblock \bibinfo{journal}{\emph{Commun. ACM}} \bibinfo{volume}{19},
  \bibinfo{number}{7} (\bibinfo{year}{1976}), \bibinfo{pages}{385--394}.
\newblock


\bibitem[Lattner and Adve(2004)]%
        {LLVM}
\bibfield{author}{\bibinfo{person}{Chris Lattner} {and}
  \bibinfo{person}{Vikram~S. Adve}.} \bibinfo{year}{2004}\natexlab{}.
\newblock \showarticletitle{LLVM: a compilation framework for lifelong program
  analysis \& transformation}.
\newblock \bibinfo{journal}{\emph{International Symposium on Code Generation
  and Optimization, 2004. CGO 2004.}} (\bibinfo{year}{2004}),
  \bibinfo{pages}{75--86}.
\newblock


\bibitem[Li et~al\mbox{.}(2022b)]%
        {CV2}
\bibfield{author}{\bibinfo{person}{Hangyu Li}, \bibinfo{person}{N. Wang},
  \bibinfo{person}{Xi Yang}, \bibinfo{person}{Xiaoyu Wang}, {and}
  \bibinfo{person}{Xinbo Gao}.} \bibinfo{year}{2022}\natexlab{b}.
\newblock \showarticletitle{Towards Semi-Supervised Deep Facial Expression
  Recognition with An Adaptive Confidence Margin}.
\newblock \bibinfo{journal}{\emph{2022 IEEE/CVF Conference on Computer Vision
  and Pattern Recognition (CVPR)}} (\bibinfo{year}{2022}),
  \bibinfo{pages}{4156--4165}.
\newblock


\bibitem[Li et~al\mbox{.}(2022a)]%
        {Li2022MEMOCM}
\bibfield{author}{\bibinfo{person}{Meiziniu Li}, \bibinfo{person}{Jialun Cao},
  \bibinfo{person}{Yongqiang Tian}, \bibinfo{person}{Tsz~On Li},
  \bibinfo{person}{Ming Wen}, {and} \bibinfo{person}{S.~C. Cheung}.}
  \bibinfo{year}{2022}\natexlab{a}.
\newblock \showarticletitle{MEMO: Coverage-guided Model Generation For Deep
  Learning Library Testing}.
\newblock \bibinfo{journal}{\emph{ArXiv}}  \bibinfo{volume}{abs/2208.01508}
  (\bibinfo{year}{2022}).
\newblock


\bibitem[Li et~al\mbox{.}(2013)]%
        {Li2013SteeringSE}
\bibfield{author}{\bibinfo{person}{You Li}, \bibinfo{person}{Zhendong Su},
  \bibinfo{person}{Linzhang Wang}, {and} \bibinfo{person}{Xuandong Li}.}
  \bibinfo{year}{2013}\natexlab{}.
\newblock \showarticletitle{Steering symbolic execution to less traveled
  paths}.
\newblock \bibinfo{journal}{\emph{Proceedings of the 2013 ACM SIGPLAN
  international conference on Object oriented programming systems languages \&
  applications}} (\bibinfo{year}{2013}).
\newblock


\bibitem[Liu et~al\mbox{.}(2021)]%
        {Liu2021DetectingMS}
\bibfield{author}{\bibinfo{person}{Dinghao Liu}, \bibinfo{person}{Qiushi Wu},
  \bibinfo{person}{Shouling Ji}, \bibinfo{person}{Kangjie Lu},
  \bibinfo{person}{Zhenguang Liu}, \bibinfo{person}{Jianhai Chen}, {and}
  \bibinfo{person}{Qinming He}.} \bibinfo{year}{2021}\natexlab{}.
\newblock \showarticletitle{Detecting Missed Security Operations Through
  Differential Checking of Object-based Similar Paths}.
\newblock \bibinfo{journal}{\emph{Proceedings of the 2021 ACM SIGSAC Conference
  on Computer and Communications Security}} (\bibinfo{year}{2021}).
\newblock


\bibitem[Liu et~al\mbox{.}(2019)]%
        {Liu2019ATOMCM}
\bibfield{author}{\bibinfo{person}{Shangqing Liu}, \bibinfo{person}{Cuiyun
  Gao}, \bibinfo{person}{Sen Chen}, \bibinfo{person}{Lun~Yiu Nie}, {and}
  \bibinfo{person}{Yang Liu}.} \bibinfo{year}{2019}\natexlab{}.
\newblock \showarticletitle{ATOM: Commit Message Generation Based on Abstract
  Syntax Tree and Hybrid Ranking}.
\newblock \bibinfo{journal}{\emph{IEEE Transactions on Software Engineering}}
  \bibinfo{volume}{48} (\bibinfo{year}{2019}), \bibinfo{pages}{1800--1817}.
\newblock


\bibitem[Luo et~al\mbox{.}(2021)]%
        {Luo2021GraphBasedFT}
\bibfield{author}{\bibinfo{person}{Wei Luo}, \bibinfo{person}{Dong Chai},
  \bibinfo{person}{Xiaoyue Run}, \bibinfo{person}{Jiang Wang},
  \bibinfo{person}{Chunrong Fang}, {and} \bibinfo{person}{Zhenyu Chen}.}
  \bibinfo{year}{2021}\natexlab{}.
\newblock \showarticletitle{Graph-Based Fuzz Testing for Deep Learning
  Inference Engines}.
\newblock \bibinfo{journal}{\emph{2021 IEEE/ACM 43rd International Conference
  on Software Engineering (ICSE)}} (\bibinfo{year}{2021}),
  \bibinfo{pages}{288--299}.
\newblock


\bibitem[MITRE(2022)]%
        {cve-mitre}
\bibfield{author}{\bibinfo{person}{MITRE}.} \bibinfo{year}{2022}\natexlab{}.
\newblock \bibinfo{title}{Common Vulnerabilities and Exposures}.
\newblock \bibinfo{howpublished}{\url{https://cve.mitre.org/index.html}}.
\newblock


\bibitem[Nejadgholi and Yang(2019)]%
        {Nejadgholi2019ASO}
\bibfield{author}{\bibinfo{person}{Mahdi Nejadgholi} {and}
  \bibinfo{person}{Jinqiu Yang}.} \bibinfo{year}{2019}\natexlab{}.
\newblock \showarticletitle{A Study of Oracle Approximations in Testing Deep
  Learning Libraries}.
\newblock \bibinfo{journal}{\emph{2019 34th IEEE/ACM International Conference
  on Automated Software Engineering (ASE)}} (\bibinfo{year}{2019}),
  \bibinfo{pages}{785--796}.
\newblock


\bibitem[Noller et~al\mbox{.}(2018)]%
        {Noller2018BadgerCA}
\bibfield{author}{\bibinfo{person}{Yannic Noller}, \bibinfo{person}{Rody
  Kersten}, {and} \bibinfo{person}{Corina~S. Pasareanu}.}
  \bibinfo{year}{2018}\natexlab{}.
\newblock \showarticletitle{Badger: complexity analysis with fuzzing and
  symbolic execution}.
\newblock \bibinfo{journal}{\emph{Proceedings of the 27th ACM SIGSOFT
  International Symposium on Software Testing and Analysis}}
  (\bibinfo{year}{2018}).
\newblock


\bibitem[Paszke et~al\mbox{.}(2019)]%
        {Paszke2019PyTorchAI}
\bibfield{author}{\bibinfo{person}{Adam Paszke}, \bibinfo{person}{Sam Gross},
  \bibinfo{person}{Francisco Massa}, \bibinfo{person}{Adam Lerer},
  \bibinfo{person}{James Bradbury}, \bibinfo{person}{Gregory Chanan},
  \bibinfo{person}{Trevor Killeen}, \bibinfo{person}{Zeming Lin},
  \bibinfo{person}{Natalia Gimelshein}, \bibinfo{person}{Luca Antiga},
  \bibinfo{person}{Alban Desmaison}, \bibinfo{person}{Andreas K{\"o}pf},
  \bibinfo{person}{Edward Yang}, \bibinfo{person}{Zach DeVito},
  \bibinfo{person}{Martin Raison}, \bibinfo{person}{Alykhan Tejani},
  \bibinfo{person}{Sasank Chilamkurthy}, \bibinfo{person}{Benoit Steiner},
  \bibinfo{person}{Lu Fang}, \bibinfo{person}{Junjie Bai}, {and}
  \bibinfo{person}{Soumith Chintala}.} \bibinfo{year}{2019}\natexlab{}.
\newblock \showarticletitle{PyTorch: An Imperative Style, High-Performance Deep
  Learning Library}.
\newblock \bibinfo{journal}{\emph{Advances in Neural Information Processing
  Systems 32: Annual Conference on Neural Information Processing Systems 2019}}
  (\bibinfo{year}{2019}).
\newblock


\bibitem[Patrick-Evans et~al\mbox{.}(2021)]%
        {PatrickEvans2021XFLNF}
\bibfield{author}{\bibinfo{person}{James Patrick-Evans},
  \bibinfo{person}{Moritz Dannehl}, {and} \bibinfo{person}{Johannes Kinder}.}
  \bibinfo{year}{2021}\natexlab{}.
\newblock \showarticletitle{XFL: Naming Functions in Binaries with Extreme
  Multi-label Learning}.
\newblock


\bibitem[Peng et~al\mbox{.}(2019)]%
        {Peng20191dVulD1}
\bibfield{author}{\bibinfo{person}{Jiaqi Peng}, \bibinfo{person}{Feng Li},
  \bibinfo{person}{Bingchang Liu}, \bibinfo{person}{Lili Xu},
  \bibinfo{person}{Binghong Liu}, \bibinfo{person}{Kai Chen}, {and}
  \bibinfo{person}{Wei Huo}.} \bibinfo{year}{2019}\natexlab{}.
\newblock \showarticletitle{1dVul: Discovering 1-Day Vulnerabilities through
  Binary Patches}.
\newblock \bibinfo{journal}{\emph{2019 49th Annual IEEE/IFIP International
  Conference on Dependable Systems and Networks (DSN)}} (\bibinfo{year}{2019}),
  \bibinfo{pages}{605--616}.
\newblock


\bibitem[Pham et~al\mbox{.}(2019)]%
        {Pham2019CRADLECV}
\bibfield{author}{\bibinfo{person}{Hung~Viet Pham}, \bibinfo{person}{Thibaud
  Lutellier}, \bibinfo{person}{Weizhen Qi}, {and} \bibinfo{person}{Lin Tan}.}
  \bibinfo{year}{2019}\natexlab{}.
\newblock \showarticletitle{CRADLE: Cross-Backend Validation to Detect and
  Localize Bugs in Deep Learning Libraries}.
\newblock \bibinfo{journal}{\emph{2019 IEEE/ACM 41st International Conference
  on Software Engineering (ICSE)}} (\bibinfo{year}{2019}),
  \bibinfo{pages}{1027--1038}.
\newblock


\bibitem[Qian et~al\mbox{.}(2022)]%
        {NLP2}
\bibfield{author}{\bibinfo{person}{Jing Qian}, \bibinfo{person}{Li Dong},
  \bibinfo{person}{Yelong Shen}, \bibinfo{person}{Furu Wei}, {and}
  \bibinfo{person}{Weizhu Chen}.} \bibinfo{year}{2022}\natexlab{}.
\newblock \showarticletitle{Controllable Natural Language Generation with
  Contrastive Prefixes}. In \bibinfo{booktitle}{\emph{Findings of the
  Association for Computational Linguistics: {ACL} 2022}}.
\newblock


\bibitem[Sen(2005)]%
        {DART}
\bibfield{author}{\bibinfo{person}{Koushik Sen}.}
  \bibinfo{year}{2005}\natexlab{}.
\newblock \showarticletitle{DART: directed automated random testing}. In
  \bibinfo{booktitle}{\emph{ACM-SIGPLAN Symposium on Programming Language
  Design and Implementation}}.
\newblock


\bibitem[Sen et~al\mbox{.}(2005)]%
        {Sen2005CUTEAC}
\bibfield{author}{\bibinfo{person}{Koushik Sen}, \bibinfo{person}{Darko
  Marinov}, {and} \bibinfo{person}{Gul~A. Agha}.}
  \bibinfo{year}{2005}\natexlab{}.
\newblock \showarticletitle{CUTE: a concolic unit testing engine for C}. In
  \bibinfo{booktitle}{\emph{ESEC/FSE-13}}.
\newblock


\bibitem[She et~al\mbox{.}(2018)]%
        {She2018NEUZZEF}
\bibfield{author}{\bibinfo{person}{Dongdong She}, \bibinfo{person}{Kexin Pei},
  \bibinfo{person}{Dave Epstein}, \bibinfo{person}{Junfeng Yang},
  \bibinfo{person}{Baishakhi Ray}, {and} \bibinfo{person}{Suman~Sekhar Jana}.}
  \bibinfo{year}{2018}\natexlab{}.
\newblock \showarticletitle{NEUZZ: Efficient Fuzzing with Neural Program
  Smoothing}.
\newblock \bibinfo{journal}{\emph{2019 IEEE Symposium on Security and Privacy
  (SP)}} (\bibinfo{year}{2018}), \bibinfo{pages}{803--817}.
\newblock


\bibitem[Singh and Khurshid(2020)]%
        {Singh2020ParallelCS}
\bibfield{author}{\bibinfo{person}{Shikhar Singh} {and}
  \bibinfo{person}{Sarfraz Khurshid}.} \bibinfo{year}{2020}\natexlab{}.
\newblock \showarticletitle{Parallel Chopped Symbolic Execution}. In
  \bibinfo{booktitle}{\emph{ICFEM}}.
\newblock


\bibitem[Stephens et~al\mbox{.}(2016)]%
        {Stephens2016DrillerAF}
\bibfield{author}{\bibinfo{person}{Nick Stephens}, \bibinfo{person}{John
  Grosen}, \bibinfo{person}{Christopher Salls}, \bibinfo{person}{Andrew
  Dutcher}, \bibinfo{person}{Ruoyu Wang}, \bibinfo{person}{Jacopo Corbetta},
  \bibinfo{person}{Yan Shoshitaishvili}, \bibinfo{person}{Christopher
  Kr{\"u}gel}, {and} \bibinfo{person}{Giovanni Vigna}.}
  \bibinfo{year}{2016}\natexlab{}.
\newblock \showarticletitle{Driller: Augmenting Fuzzing Through Selective
  Symbolic Execution}. In \bibinfo{booktitle}{\emph{NDSS}}.
\newblock


\bibitem[Sui and Xue(2016)]%
        {SVF}
\bibfield{author}{\bibinfo{person}{Yulei Sui} {and} \bibinfo{person}{Jingling
  Xue}.} \bibinfo{year}{2016}\natexlab{}.
\newblock \showarticletitle{SVF: interprocedural static value-flow analysis in
  LLVM}.
\newblock \bibinfo{journal}{\emph{Proceedings of the 25th International
  Conference on Compiler Construction}} (\bibinfo{year}{2016}).
\newblock


\bibitem[Trabish et~al\mbox{.}(2018)]%
        {Trabish2018ChoppedSE}
\bibfield{author}{\bibinfo{person}{David Trabish}, \bibinfo{person}{Andrea
  Mattavelli}, \bibinfo{person}{Noam Rinetzky}, {and} \bibinfo{person}{Cristian
  Cadar}.} \bibinfo{year}{2018}\natexlab{}.
\newblock \showarticletitle{Chopped Symbolic Execution}.
\newblock \bibinfo{journal}{\emph{2018 IEEE/ACM 40th International Conference
  on Software Engineering (ICSE)}} (\bibinfo{year}{2018}),
  \bibinfo{pages}{350--360}.
\newblock


\bibitem[Wang et~al\mbox{.}(2017a)]%
        {Wang2017SkyfireDS}
\bibfield{author}{\bibinfo{person}{Junjie Wang}, \bibinfo{person}{Bihuan Chen},
  \bibinfo{person}{Lei Wei}, {and} \bibinfo{person}{Yang Liu}.}
  \bibinfo{year}{2017}\natexlab{a}.
\newblock \showarticletitle{Skyfire: Data-Driven Seed Generation for Fuzzing}.
\newblock \bibinfo{journal}{\emph{2017 IEEE Symposium on Security and Privacy
  (SP)}} (\bibinfo{year}{2017}), \bibinfo{pages}{579--594}.
\newblock


\bibitem[Wang et~al\mbox{.}(2022)]%
        {Wang2022EAGLECE}
\bibfield{author}{\bibinfo{person}{Jiannan Wang}, \bibinfo{person}{Thibaud
  Lutellier}, \bibinfo{person}{Shangshu Qian}, \bibinfo{person}{Hung~Viet
  Pham}, {and} \bibinfo{person}{Lin Tan}.} \bibinfo{year}{2022}\natexlab{}.
\newblock \showarticletitle{EAGLE: Creating Equivalent Graphs to Test Deep
  Learning Libraries}.
\newblock \bibinfo{journal}{\emph{2022 IEEE/ACM 44th International Conference
  on Software Engineering (ICSE)}} (\bibinfo{year}{2022}),
  \bibinfo{pages}{798--810}.
\newblock


\bibitem[Wang et~al\mbox{.}(2017b)]%
        {Wang2017AutomaticDA}
\bibfield{author}{\bibinfo{person}{Yu Wang}, \bibinfo{person}{Linzhang Wang},
  \bibinfo{person}{Tingting Yu}, \bibinfo{person}{Jianhua Zhao}, {and}
  \bibinfo{person}{Xuandong Li}.} \bibinfo{year}{2017}\natexlab{b}.
\newblock \showarticletitle{Automatic detection and validation of race
  conditions in interrupt-driven embedded software}.
\newblock \bibinfo{journal}{\emph{Proceedings of the 26th ACM SIGSOFT
  International Symposium on Software Testing and Analysis}}
  (\bibinfo{year}{2017}).
\newblock


\bibitem[Wang et~al\mbox{.}(2020)]%
        {Wang2020DeepLL}
\bibfield{author}{\bibinfo{person}{Zan Wang}, \bibinfo{person}{Ming Yan},
  \bibinfo{person}{Junjie Chen}, \bibinfo{person}{Shuang Liu}, {and}
  \bibinfo{person}{Dongdi Zhang}.} \bibinfo{year}{2020}\natexlab{}.
\newblock \showarticletitle{Deep learning library testing via effective model
  generation}.
\newblock \bibinfo{journal}{\emph{Proceedings of the 28th ACM Joint Meeting on
  European Software Engineering Conference and Symposium on the Foundations of
  Software Engineering}} (\bibinfo{year}{2020}).
\newblock


\bibitem[Wei et~al\mbox{.}(2022)]%
        {Wei2022FreeLF}
\bibfield{author}{\bibinfo{person}{Anjiang Wei}, \bibinfo{person}{Y. Deng},
  \bibinfo{person}{Chenyuan Yang}, {and} \bibinfo{person}{Lingming Zhang}.}
  \bibinfo{year}{2022}\natexlab{}.
\newblock \showarticletitle{Free Lunch for Testing: Fuzzing Deep-Learning
  Libraries from Open Source}.
\newblock \bibinfo{journal}{\emph{2022 IEEE/ACM 44th International Conference
  on Software Engineering (ICSE)}} (\bibinfo{year}{2022}),
  \bibinfo{pages}{995--1007}.
\newblock


\bibitem[Wu et~al\mbox{.}(2022)]%
        {Wu2022DeepCovCG}
\bibfield{author}{\bibinfo{person}{Jiawei Wu}, \bibinfo{person}{Senyi Li},
  \bibinfo{person}{Junqiang Li}, \bibinfo{person}{Long Luo},
  \bibinfo{person}{Hongfang Yu}, {and} \bibinfo{person}{Gang Sun}.}
  \bibinfo{year}{2022}\natexlab{}.
\newblock \showarticletitle{DeepCov: Coverage Guided Deep Learning Framework
  Fuzzing}.
\newblock \bibinfo{journal}{\emph{2022 7th IEEE International Conference on
  Data Science in Cyberspace (DSC)}} (\bibinfo{year}{2022}),
  \bibinfo{pages}{399--404}.
\newblock


\bibitem[Wu et~al\mbox{.}(2018)]%
        {Wu2018FUZETF}
\bibfield{author}{\bibinfo{person}{Wei Wu}, \bibinfo{person}{Yueqi Chen},
  \bibinfo{person}{Jun Xu}, \bibinfo{person}{Xinyu Xing},
  \bibinfo{person}{Xiaorui Gong}, {and} \bibinfo{person}{Wei Zou}.}
  \bibinfo{year}{2018}\natexlab{}.
\newblock \showarticletitle{FUZE: Towards Facilitating Exploit Generation for
  Kernel Use-After-Free Vulnerabilities}. In \bibinfo{booktitle}{\emph{USENIX
  Security Symposium}}.
\newblock


\bibitem[Xiao et~al\mbox{.}(2020)]%
        {Xiao2020MVPDV}
\bibfield{author}{\bibinfo{person}{Yang Xiao}, \bibinfo{person}{Bihuan Chen},
  \bibinfo{person}{Chendong Yu}, \bibinfo{person}{Zhengzi Xu},
  \bibinfo{person}{Zimu Yuan}, \bibinfo{person}{Feng Li},
  \bibinfo{person}{Binghong Liu}, \bibinfo{person}{Yang Liu},
  \bibinfo{person}{Wei Huo}, \bibinfo{person}{Wei Zou}, {and}
  \bibinfo{person}{Wenchang Shi}.} \bibinfo{year}{2020}\natexlab{}.
\newblock \showarticletitle{MVP: Detecting Vulnerabilities using Patch-Enhanced
  Vulnerability Signatures}. In \bibinfo{booktitle}{\emph{USENIX Security
  Symposium}}.
\newblock


\bibitem[Xie et~al\mbox{.}(2022)]%
        {DocTer}
\bibfield{author}{\bibinfo{person}{Danning Xie}, \bibinfo{person}{Yitong Li},
  \bibinfo{person}{Mijung Kim}, \bibinfo{person}{Hung~Viet Pham},
  \bibinfo{person}{Lin Tan}, \bibinfo{person}{X. Zhang}, {and}
  \bibinfo{person}{Michael~W. Godfrey}.} \bibinfo{year}{2022}\natexlab{}.
\newblock \showarticletitle{DocTer: documentation-guided fuzzing for testing
  deep learning API functions}.
\newblock \bibinfo{journal}{\emph{Proceedings of the 31st ACM SIGSOFT
  International Symposium on Software Testing and Analysis}}
  (\bibinfo{year}{2022}).
\newblock


\bibitem[Yang et~al\mbox{.}(2023)]%
        {Yang2023FuzzingAD}
\bibfield{author}{\bibinfo{person}{Chenyuan Yang}, \bibinfo{person}{Yinlin
  Deng}, \bibinfo{person}{Jiayi Yao}, \bibinfo{person}{Yuxing Tu},
  \bibinfo{person}{Hanchi Li}, {and} \bibinfo{person}{Lingming Zhang}.}
  \bibinfo{year}{2023}\natexlab{}.
\newblock \showarticletitle{Fuzzing Automatic Differentiation in Deep-Learning
  Libraries}.
\newblock \bibinfo{journal}{\emph{ArXiv}}  \bibinfo{volume}{abs/2302.04351}
  (\bibinfo{year}{2023}).
\newblock


\bibitem[Zhang et~al\mbox{.}(2021a)]%
        {Duo}
\bibfield{author}{\bibinfo{person}{Xufan Zhang}, \bibinfo{person}{Jiawei Liu},
  \bibinfo{person}{Ning Sun}, \bibinfo{person}{Chunrong Fang},
  \bibinfo{person}{Jia Liu}, \bibinfo{person}{Jiang Wang},
  \bibinfo{person}{Dong Chai}, {and} \bibinfo{person}{Zhenyu Chen}.}
  \bibinfo{year}{2021}\natexlab{a}.
\newblock \showarticletitle{Duo: Differential Fuzzing for Deep Learning
  Operators}.
\newblock \bibinfo{journal}{\emph{IEEE Transactions on Reliability}}
  \bibinfo{volume}{70} (\bibinfo{year}{2021}), \bibinfo{pages}{1671--1685}.
\newblock


\bibitem[Zhang et~al\mbox{.}(2021b)]%
        {Predoo}
\bibfield{author}{\bibinfo{person}{Xufan Zhang}, \bibinfo{person}{Ning Sun},
  \bibinfo{person}{Chunrong Fang}, \bibinfo{person}{Jiawei Liu},
  \bibinfo{person}{Jia Liu}, \bibinfo{person}{Dong Chai},
  \bibinfo{person}{Jiang Wang}, {and} \bibinfo{person}{Zhenyu Chen}.}
  \bibinfo{year}{2021}\natexlab{b}.
\newblock \showarticletitle{Predoo: precision testing of deep learning
  operators}.
\newblock \bibinfo{journal}{\emph{Proceedings of the 30th ACM SIGSOFT
  International Symposium on Software Testing and Analysis}}
  (\bibinfo{year}{2021}).
\newblock


\bibitem[Zhang et~al\mbox{.}(2020)]%
        {Zhang2020MultiplexSE}
\bibfield{author}{\bibinfo{person}{Yufeng Zhang}, \bibinfo{person}{Zhenbang
  Chen}, \bibinfo{person}{Ziqi Shuai}, \bibinfo{person}{Tianqi Zhang},
  \bibinfo{person}{Kenli Li}, {and} \bibinfo{person}{Ji Wang}.}
  \bibinfo{year}{2020}\natexlab{}.
\newblock \showarticletitle{Multiplex Symbolic Execution: Exploring Multiple
  Paths by Solving Once}.
\newblock \bibinfo{journal}{\emph{2020 35th IEEE/ACM International Conference
  on Automated Software Engineering (ASE)}} (\bibinfo{year}{2020}),
  \bibinfo{pages}{846--857}.
\newblock


\end{thebibliography}




\end{document}